\documentclass[%
 aip,
 amsmath,amssymb,
 reprint,%
]{revtex4-1}

\usepackage{graphicx}% Include figure files
\usepackage{dcolumn}% Align table columns on decimal point
\usepackage{bm}% bold math
\usepackage[utf8]{inputenc}
\usepackage[T1]{fontenc}
\usepackage{mathptmx}
\usepackage{etoolbox}

\makeatletter
\def\@email#1#2{%
 \endgroup
 \patchcmd{\titleblock@produce}
  {\frontmatter@RRAPformat}
  {\frontmatter@RRAPformat{\produce@RRAP{*#1\href{mailto:#2}{#2}}}\frontmatter@RRAPformat}
  {}{}
}%
\makeatother

\begin{document}

\preprint{AIP/123-QED}

\title{Development of the Self-Modulation Instability of a Relativistic Proton Bunch in Plasma}
% Force line breaks with \\
\author{L.~Verra}
\email{livio.verra@cern.ch}
\affiliation{CERN, 1211 Geneva 23, Switzerland}
\author{S.~Wyler}
\affiliation{Ecole Polytechnique Federale de Lausanne (EPFL), Swiss Plasma Center (SPC), 1015 Lausanne, Switzerland}
\author{T.~Nechaeva}
\affiliation{Max Planck Institute for Physics, 80805 Munich, Germany}
\author{J.~Pucek}
\affiliation{Max Planck Institute for Physics, 80805 Munich, Germany}
\author{V.~Bencini}
\affiliation{CERN, 1211 Geneva 23, Switzerland}
\affiliation{John Adams Institute, Oxford University, Oxford OX1 3RH, United Kingdom}
\author{M.~Bergamaschi} 
\affiliation{Max Planck Institute for Physics, 80805 Munich, Germany}
\author{L.~Ranc}
\affiliation{Max Planck Institute for Physics, 80805 Munich, Germany}
\author{G.~Zevi Della Porta}
\affiliation{CERN, 1211 Geneva 23, Switzerland}
\affiliation{Max Planck Institute for Physics, 80805 Munich, Germany}
\author{E.~Gschwendtner}
\affiliation{CERN, 1211 Geneva 23, Switzerland}
\author{P.~Muggli}
\affiliation{Max Planck Institute for Physics, 80805 Munich, Germany}
\collaboration{AWAKE Collaboration}
\noaffiliation
\author{R.~Agnello}
\affiliation{Ecole Polytechnique Federale de Lausanne (EPFL), Swiss Plasma Center (SPC), 1015 Lausanne, Switzerland}
\author{C.C.~Ahdida}
\affiliation{CERN, 1211 Geneva 23, Switzerland}
\author{C.~Amoedo}
\affiliation{CERN, 1211 Geneva 23, Switzerland}
\author{Y.~Andrebe}
\affiliation{Ecole Polytechnique Federale de Lausanne (EPFL), Swiss Plasma Center (SPC), 1015 Lausanne, Switzerland}
\author{O.~Apsimon}
\affiliation{University of Manchester M13 9PL, Manchester M13 9PL, United Kingdom}
\affiliation{Cockcroft Institute, Warrington WA4 4AD, United Kingdom}
\author{R.~Apsimon}
\affiliation{Cockcroft Institute, Warrington WA4 4AD, United Kingdom} % left as in previous papers
\affiliation{Lancaster University, Lancaster LA1 4YB, United Kingdom}
\author{J.M.~Arnesano}
\affiliation{CERN, 1211 Geneva 23, Switzerland}
\author{P.~Blanchard}
\affiliation{Ecole Polytechnique Federale de Lausanne (EPFL), Swiss Plasma Center (SPC), 1015 Lausanne, Switzerland}
\author{P.N.~Burrows}
\affiliation{John Adams Institute, Oxford University, Oxford OX1 3RH, United Kingdom}
\author{B.~Buttensch{\"o}n}
\affiliation{Max Planck Institute for Plasma Physics, 17491 Greifswald, Germany}
\author{A.~Caldwell}
\affiliation{Max Planck Institute for Physics, 80805 Munich, Germany}
\author{M.~Chung}
\affiliation{UNIST, Ulsan 44919, Republic of Korea}
\author{D.A.~Cooke}
\affiliation{UCL, London WC1 6BT, United Kingdom}
\author{C.~Davut}
\affiliation{University of Manchester M13 9PL, Manchester M13 9PL, United Kingdom}
\affiliation{Cockcroft Institute, Warrington WA4 4AD, United Kingdom} % left as in previous papers
\author{G.~Demeter}
\affiliation{Wigner Research Centre for Physics, 1121 Budapest, Hungary}
\author{A.C.~Dexter}
\affiliation{Cockcroft Institute, Warrington WA4 4AD, United Kingdom} % left as in previous papers
\affiliation{Lancaster University, Lancaster LA1 4YB, United Kingdom}
\author{S.~Doebert}
\affiliation{CERN, 1211 Geneva 23, Switzerland}
\author{F.A.~Elverson}
\affiliation{CERN, 1211 Geneva 23, Switzerland}
\author{J.~Farmer}
\affiliation{Max Planck Institute for Physics, 80805 Munich, Germany}
\author{A.~Fasoli}
\affiliation{Ecole Polytechnique Federale de Lausanne (EPFL), Swiss Plasma Center (SPC), 1015 Lausanne, Switzerland}
\author{R.~Fonseca}
\affiliation{ISCTE - Instituto Universit\'{e}ario de Lisboa, 1049-001 Lisbon, Portugal}  % left as in previous papers
\affiliation{GoLP/Instituto de Plasmas e Fus\~{a}o Nuclear, Instituto Superior T\'{e}cnico, Universidade de Lisboa, 1049-001 Lisbon, Portugal}
\author{I.~Furno}
\affiliation{Ecole Polytechnique Federale de Lausanne (EPFL), Swiss Plasma Center (SPC), 1015 Lausanne, Switzerland}
\author{A.~Gorn}
\affiliation{Budker Institute of Nuclear Physics SB RAS, 630090 Novosibirsk, Russia}
\affiliation{Novosibirsk State University, 630090 Novosibirsk , Russia}
\author{E.~Granados}
\affiliation{CERN, 1211 Geneva 23, Switzerland}
\author{M.~Granetzny}
\affiliation{University of Wisconsin, Madison, WI 53706, USA}
\author{T.~Graubner}
\affiliation{Philipps-Universit{\"a}t Marburg, 35032 Marburg, Germany}
\author{O.~Grulke}
\affiliation{Max Planck Institute for Plasma Physics, 17491 Greifswald, Germany}
\affiliation{Technical University of Denmark, 2800 Kgs. Lyngby, Denmark}
\author{E.~Guran}
\affiliation{CERN, 1211 Geneva 23, Switzerland}
\author{J.~Henderson}
\affiliation{Cockcroft Institute, Warrington WA4 4AD, United Kingdom}
\affiliation{STFC/ASTeC, Daresbury Laboratory, Warrington WA4 4AD, United Kingdom}
\author{M.Á.~Kedves}
\affiliation{Wigner Research Centre for Physics, 1121 Budapest, Hungary}
\author{S.-Y.~Kim}
\affiliation{UNIST, Ulsan 44919, Republic of Korea}
\affiliation{CERN, 1211 Geneva 23, Switzerland}
\author{F.~Kraus}
\affiliation{Philipps-Universit{\"a}t Marburg, 35032 Marburg, Germany}
\author{M.~Krupa}
\affiliation{CERN, 1211 Geneva 23, Switzerland}
\author{T.~Lefevre}
\affiliation{CERN, 1211 Geneva 23, Switzerland}
\author{L.~Liang}
\affiliation{University of Manchester M13 9PL, Manchester M13 9PL, United Kingdom}
\affiliation{Cockcroft Institute, Warrington WA4 4AD, United Kingdom}
\author{S.~Liu}
\affiliation{TRIUMF, Vancouver, BC V6T 2A3, Canada}
\author{N.~Lopes}
\affiliation{GoLP/Instituto de Plasmas e Fus\~{a}o Nuclear, Instituto Superior T\'{e}cnico, Universidade de Lisboa, 1049-001 Lisbon, Portugal}
\author{K.~Lotov}
\affiliation{Budker Institute of Nuclear Physics SB RAS, 630090 Novosibirsk, Russia}
\affiliation{Novosibirsk State University, 630090 Novosibirsk , Russia}
\author{M.~Martinez~Calderon}
\affiliation{CERN, 1211 Geneva 23, Switzerland}
\author{S.~Mazzoni}
\affiliation{CERN, 1211 Geneva 23, Switzerland}
\author{K.~Moon}
\affiliation{UNIST, Ulsan 44919, Republic of Korea}
\author{P.I.~Morales~Guzm\'{a}n}
\affiliation{Max Planck Institute for Physics, 80805 Munich, Germany}
\author{M.~Moreira}
\affiliation{GoLP/Instituto de Plasmas e Fus\~{a}o Nuclear, Instituto Superior T\'{e}cnico, Universidade de Lisboa, 1049-001 Lisbon, Portugal}
\author{C.~Pakuza}
\affiliation{John Adams Institute, Oxford University, Oxford OX1 3RH, United Kingdom}
\author{F.~Pannell}
\affiliation{UCL, London WC1 6BT, United Kingdom}
\author{A.~Pardons}
\affiliation{CERN, 1211 Geneva 23, Switzerland}
\author{K.~Pepitone}
\affiliation{Angstrom Laboratory, Department of Physics and Astronomy, 752 37 Uppsala, Sweden}
\author{E.~Poimendidou}
\affiliation{CERN, 1211 Geneva 23, Switzerland}
\author{A.~Pukhov}
\affiliation{Heinrich-Heine-Universit{\"a}t D{\"u}sseldorf, 40225 D{\"u}sseldorf, Germany}
\author{R.L.~Ramjiawan}
\affiliation{CERN, 1211 Geneva 23, Switzerland}
\affiliation{John Adams Institute, Oxford University, Oxford OX1 3RH, United Kingdom}
\author{S.~Rey}
\affiliation{CERN, 1211 Geneva 23, Switzerland}
\author{R.~Rossel}
\affiliation{CERN, 1211 Geneva 23, Switzerland}
\author{H.~Saberi}
\affiliation{University of Manchester M13 9PL, Manchester M13 9PL, United Kingdom}
\affiliation{Cockcroft Institute, Warrington WA4 4AD, United Kingdom}
\author{O.~Schmitz}
\affiliation{University of Wisconsin, Madison, WI 53706, USA}
\author{E.~Senes}
\affiliation{CERN, 1211 Geneva 23, Switzerland}
\author{F.~Silva}
\affiliation{INESC-ID, Instituto Superior Técnico, Universidade de Lisboa, 1049-001 Lisbon, Portugal}
\author{L.~Silva}
\affiliation{GoLP/Instituto de Plasmas e Fus\~{a}o Nuclear, Instituto Superior T\'{e}cnico, Universidade de Lisboa, 1049-001 Lisbon, Portugal}
\author{B.~Spear}
\affiliation{John Adams Institute, Oxford University, Oxford OX1 3RH, United Kingdom}
\author{C.~Stollberg}
\affiliation{Ecole Polytechnique Federale de Lausanne (EPFL), Swiss Plasma Center (SPC), 1015 Lausanne, Switzerland}
\author{A.~Sublet}
\affiliation{CERN, 1211 Geneva 23, Switzerland}
\author{C.~Swain}
\affiliation{Cockcroft Institute, Warrington WA4 4AD, United Kingdom}
\affiliation{University of Liverpool, Liverpool L69 7ZE, United Kingdom}
\author{A.~Topaloudis}
\affiliation{CERN, 1211 Geneva 23, Switzerland}
\author{N.~Torrado}
\affiliation{GoLP/Instituto de Plasmas e Fus\~{a}o Nuclear, Instituto Superior T\'{e}cnico, Universidade de Lisboa, 1049-001 Lisbon, Portugal}
\affiliation{CERN, 1211 Geneva 23, Switzerland}
\author{P.~Tuev}
\affiliation{Budker Institute of Nuclear Physics SB RAS, 630090 Novosibirsk, Russia}
\affiliation{Novosibirsk State University, 630090 Novosibirsk , Russia}
\author{M.~Turner}
\affiliation{CERN, 1211 Geneva 23, Switzerland}
\author{F.~Velotti}
\affiliation{CERN, 1211 Geneva 23, Switzerland}
\author{V.~Verzilov}
\affiliation{TRIUMF, Vancouver, BC V6T 2A3, Canada}
\author{J.~Vieira}
\affiliation{GoLP/Instituto de Plasmas e Fus\~{a}o Nuclear, Instituto Superior T\'{e}cnico, Universidade de Lisboa, 1049-001 Lisbon, Portugal}
\author{M.~Weidl}
\affiliation{Max Planck Institute for Plasma Physics, 80805 Garching, Germany}
\author{C.~Welsch}
\affiliation{Cockcroft Institute, Warrington WA4 4AD, United Kingdom}
\affiliation{University of Liverpool, Liverpool L69 7ZE, United Kingdom}
\author{M.~Wendt}
\affiliation{CERN, 1211 Geneva 23, Switzerland}
\author{M.~Wing}
\affiliation{UCL, London WC1 6BT, United Kingdom}
\author{J.~Wolfenden}
\affiliation{Cockcroft Institute, Warrington WA4 4AD, United Kingdom}
\affiliation{University of Liverpool, Liverpool L69 7ZE, United Kingdom}
\author{B.~Woolley}
\affiliation{CERN, 1211 Geneva 23, Switzerland}
\author{G.~Xia}
\affiliation{Cockcroft Institute, Warrington WA4 4AD, United Kingdom}
\affiliation{University of Manchester M13 9PL, Manchester M13 9PL, United Kingdom}
\author{V.~Yarygova}
\affiliation{Budker Institute of Nuclear Physics SB RAS, 630090 Novosibirsk, Russia}
\affiliation{Novosibirsk State University, 630090 Novosibirsk , Russia}
\author{M.~Zepp}
\affiliation{University of Wisconsin, Madison, WI 53706, USA}

\date{\today}% It is always \today, today,
             %  but any date may be explicitly specified

\begin{abstract}
Self-modulation is a beam-plasma instability that is useful to drive large-amplitude wakefields with bunches much longer than the plasma skin depth.
We present experimental results showing that, when increasing the ratio between the initial transverse size of the bunch and the plasma skin depth, the instability occurs later along the bunch, or not at all, over a fixed plasma length, because the amplitude of the initial wakefields decreases.
We show cases for which self-modulation does not develop and we introduce a simple model discussing the conditions for which it would not occur after any plasma length.
Changing bunch size and plasma electron density also changes the growth rate of the instability.
We discuss the impact of these results on the design of a particle accelerator based on the self-modulation instability seeded by a relativistic ionization front, such as the future upgrade of the AWAKE experiment.
\end{abstract}

\maketitle

\section{Introduction}

The self-modulation instability (SMI)~\cite{KUMAR:GROWTH,KARL:PRL,MARLENE:PRL} is a beam-plasma instability that can develop when a relativistic charged particle bunch, propagating in a plasma with density $n_{pe}$, has root mean square (rms) length $\sigma_z$ much longer than the plasma skin depth $c/\omega_{pe}$, where  $c$ is the speed of light and $\omega_{pe}=\sqrt{\frac{n_{pe}e^2}{m_e \varepsilon_0}}$ is the plasma electron angular frequency ($m_e$ and $e$ are the mass and charge of the electron, $\varepsilon_0$ is the vacuum permittivity).
When the bunch enters a preformed plasma, the features and imperfections of the initial distribution of the bunch density~\cite{FABIAN:PRL,KOSTANTIN:NOISE} drive initial wakefields~\cite{PWFA:CHEN} that act on the bunch itself.
The transverse component of these wakefields generates a periodic focusing and defocusing force, 
%initially 
 modulating the radius of the bunch. 
Since the longitudinal relative motion of highly relativistic particles due to the action of longitudinal component of the wakefields is negligible, the modulation of the radius results in a longitudinal modulation of the bunch density.

\par The instability grows exponentially from the initial modulation, and microbunches form on axis when the effect of the focusing force becomes significant when compared to the natural divergence of the bunch. 
According to linear theory~\cite{KUMAR:GROWTH,PUKHOV:GROWTH,SCH:GROWTH}, the amplitude of the transverse wakefields $W_{\perp}$ grows along the plasma ($z$) and along the bunch ($\xi$), starting from the initial amplitude $W_{\perp 0}(\xi)$, as $W_{\perp}(z,\xi) = W_{\perp 0}(\xi) \exp(\Gamma(z,\xi) z)$, where $\Gamma$ is the growth rate of the instability.
% Hence, depending on the relative amplitude of the two terms, $W_{\perp}$ and the modulation structure are dominated by the amplitude of the initial wakefields $W_{\perp0}$ or by the growth rate of the instability $\Gamma$.
As the instability grows along the bunch, the full modulation into microbunches first occurs late along the Gaussian bunch, and progressively reaches its front during propagation in plasma.
Thus, at the front the effect of $\Gamma$ is smaller than at the back~\cite{KUMAR:GROWTH,PUKHOV:GROWTH,SCH:GROWTH} ($\Gamma\propto \xi^{1/3}$).
% and the effect of only $W_{\perp 0}$ remains dominant on the bunch evolution for a longer distance in plasma.
% Thus, the depth of the modulation measured at the bunch front is a probe for $W_{\perp 0}$, that causes the initial radial modulation.
% Thus, the earliest time along the bunch where SMI is observed is a probe for $W_{\perp 0}$, that causes the initial radial modulation.

Once the bunch is fully modulated into a train of microbunches (i.e., SMI reaches saturation), it resonantly drives wakefields that can be used for high-gradient particle acceleration~\cite{AW:NATURE}.
The development of SMI may be ensured and made reproducible by seeding~\cite{FABIAN:PRL,LIVIO:PRL, FANG:PhysRevLett.112.045001} or may be suppressed by decreasing the amplitude of the initial wakefields.

\par In the context of the Advanced WAKefield Experiment (AWAKE) at CERN~\cite{PATRIC:READINESS}, we previously demonstrated that a long, Gaussian, relativistic proton ($p^+$) bunch undergoes the self-modulation process in plasma~\cite{KARL:PRL,MARLENE:PRL} and that externally injected electrons can be accelerated to GeV-energies~\cite{AW:NATURE}.
We also showed that self-modulation can be seeded by a relativistic ionization front (RIF) copropagating within the $p^+$ bunch~\cite{FABIAN:PRL}, or by wakefields driven by a preceding, short electron bunch~\cite{LIVIO:PRL}.
In case of RIF seeding, seed wakefields are provided by the fast onset of the beam-plasma interaction at the location of the RIF.
% A transition from the instability to the seeded regime was observed when the local bunch density at the RIF location overcomes a threshold value~\cite{FABIAN:PRL}.
\par The possibility of varying the initial parameters of the bunch and plasma allows for further studies on the development of SMI.
In previous experiments, we showed that the growth rate of the instability increases when increasing $n_{pe}$~\cite{MARLENE:PRL} or when increasing the peak bunch charge density $n_{b0}$~\cite{LIVIO:PRL}.
It is also interesting to measure the 
%(to date, unknown) 
dependence of the amplitude of the initial transverse wakefields and of the development of the instability on initial parameters.
When $W_{\perp 0}$ is decreased, it takes a longer propagation distance in plasma for the initial modulation to form, and therefore for SMI to grow.
Thus, the earliest time along the bunch where SMI is observed is a probe for $W_{\perp 0}$, that causes the initial modulation.

\par We perform experiments in which the bunch density is much lower than the plasma density ($n_{b0}/n_{pe}<0.08$ in all cases), and therefore initial wakefields are in the linear regime.
The amplitude of the initial transverse wakefields is proportional to $n_{b0}$ and inversely proportional to the ratio between the initial rms transverse size of the bunch~\cite{JONES:LIN_THEORY} $\sigma_{r0}$ and $c/\omega_{pe}$ ($W_{\perp0}\propto n_{b0}( c/\omega_{pe})\propto(1/\sigma_{r0}^2)( c/\omega_{pe})$), as the wakefields are defined by the amount of bunch charge contained within $c/\omega_{pe}$.
% As the initial modulation is caused by $W_{\perp 0}$, the depth of the modulation depends on the amplitude of the wakefields initially driven by the bunch.
Hence, when increasing $\sigma_{r0}$ (keeping constant the charge and the other parameters of the bunch), both $W_{\perp 0}$ and $\Gamma$ decrease~\cite{KUMAR:GROWTH, PUKHOV:GROWTH, SCH:GROWTH} ($\Gamma \propto \omega_{pe}(\xi n_{b0}/n_{pe})^{(1/3)}$), resulting in a smaller amplitude of the wakefields during the growth of SMI, and eventually in the suppression of the development of the instability.
% over a given propagation distance in plasma.
When increasing $n_{pe}$, $W_{\perp 0}$ decreases, whereas $\Gamma$ increases~\cite{KUMAR:GROWTH,PUKHOV:GROWTH,SCH:GROWTH,MARLENE:PRL}.
% , and previous experimental results~\cite{MARLENE:PRL} showed that the effect on $\Gamma$ causes an increase of the overall amplitude of the wakefields during the growth, when increasing $n_{pe}$.
% However, as the instability grows along the bunch, over a fixed length of the plasma it may not reach saturation at the front of a Gaussian bunch. 
% At the front of the bunch, the effect of $\Gamma$ is smaller than at the back and the effect of $W_{\perp 0}$ remains dominant for a longer distance in plasma.

\par In this paper, we present experimental results showing that, when increasing $\sigma_{r0}$ or $n_{pe}$, the modulation becomes observable only later along the bunch, due to a lower $W_{\perp 0}$, independently of $\Gamma$.
% and of the amplitude of the wakefields during growth.
We perform the experiments by measuring the charge density distribution of the bunch on time-resolved images obtained after propagation in plasma.
The occurrence of SMI may be directly visible from time-resolved images and, with higher sensitivity, from the power spectrum of the discrete Fourier transform of their on-axis profile.
% We calculate the power spectrum of the discrete Fourier transform of their on-axis profiles to detect the modulation and measure its frequency.
In particular, we show that the development of SMI can be controlled, delayed, and even suppressed,
% over a fixed plasma length, 
by varying these initial parameters.
We also discuss the implications for the design of a plasma wakefield accelerator based on the self-modulation of the drive bunch, such as the future upgrade of AWAKE.

% \par This is an important result for AWAKE because it shows that the development of SMI can be controlled, and even suppressed, over a fixed plasma length, by tailoring $\sigma_r$ and  $n_{pe}$.
% In particular, we show that a bunch with large transverse size reaches only late along the bunch a modulation depth deeper than that due to the random variation in the incoming charge density distribution.

\section{Experimental Setup}
\begin{figure}[h!]
\centering
\includegraphics[scale=0.55]{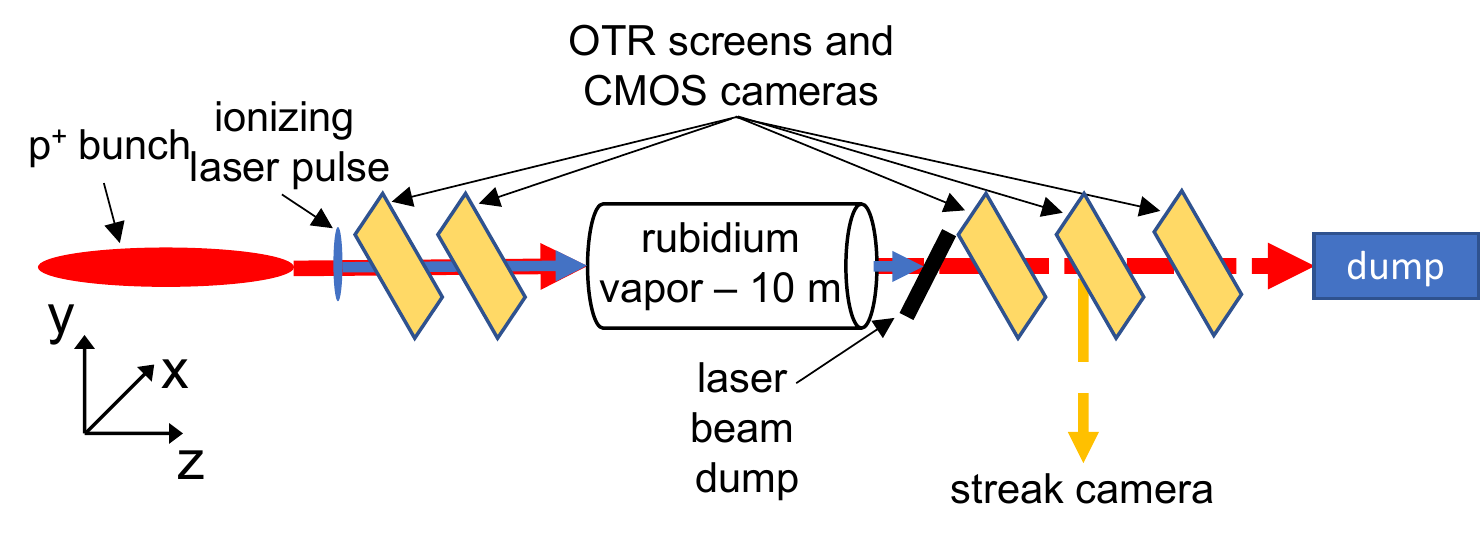}
\caption{
Schematic of the AWAKE experimental setup: the ionizing laser pulse (blue) enters the 10-m-long vapor source (white cylinder) ahead of the $p^+$ bunch (red) and singly ionizes the rubidium atoms, creating the plasma.
The optical transition radiation produced by aluminum-coated screens (yellow trapezoids) when the bunch enters them is imaged on the chip of CMOS cameras, producing time-integrated images (e.g., Fig.~\ref{fig:2}(a) and (b)), or on the entrance slit of a streak camera, producing time-resolved images (e.g., Fig.~\ref{fig:3}). 
The two screens upstream of the vapor source are extracted from the beamline during the experiments with plasma.
}
\label{fig:1}
\end{figure}

In AWAKE (Fig.~\ref{fig:1}), the plasma is generated by ionizing rubidium vapor contained in a $10$-m-long source with a $\sim 120\,$fs, $\sim 100\,$mJ laser pulse focused to a radius of $\sim 1\,$mm~\cite{GABOR:PLASMA}.
The vapor density $n_{vap}$ is controlled by varying the temperature of the reservoirs containing the rubidium and of the vapor source~\cite{SOURCE:Plyushchev_2018}.
Previous experimental results showed that the laser pulse singly ionizes all the rubidium atoms on its path~\cite{KARL:PRL}, thus, $n_{pe}=n_{vap}$.
In the experiments presented here, we use $n_{pe}=0.97\times 10^{14}\,$cm$^{-3}$ and $n_{pe}=7.3\times 10^{14}\,$cm$^{-3}$, which in the following we refer to as low and high plasma electron densities, respectively. 
The CERN Super Proton Synchrotron (SPS) delivers the 44.9\,nC, $\sigma_z = 6.3\,$cm bunch of $400\,$GeV/c protons~\cite{SCHMIDT:PROTON_LINE}.
The final focusing system of the transport line provides flexibility for placing the waist of the beam at different locations $z^{*}$ with respect to the plasma entrance ($z=0$), and for varying the transverse size $\sigma_{r0}$ of the bunch entering the plasma.
By changing $n_{pe}$ and $\sigma_{r0}$, we therefore vary the ratio $\sigma_{r0}/(c/\omega_{pe})\propto\sigma_{r0}\sqrt{n_{pe}}$, and thus the amplitude of the initial wakefields as well as the growth rate.
% that is a key parameter for wakefield generation and, thus, for the development of the self-modulation process.
%%%%%%%%%
%%%%%%%%%%%%%%%%%%%%%%%%
\par The transverse size of the beam at different locations along the beamline is measured on screens (yellow trapezoids in Fig.~\ref{fig:1}) emitting optical transition radiation (OTR) when the bunch traverses them. 
The OTR is transported and imaged onto the chip of CMOS cameras, producing time-integrated images of the $p^+$ bunch transverse distribution.
The light produced by one of these screens, positioned $3.5\,$m downstream of the plasma exit ($z=13.5\,$m), is also sent to a streak camera, producing a time-resolved image of the $p^+$ bunch charge density distribution in a 180-$\mu$m-wide slice (the spatial resolution of the optical system~\cite{Nechaeva_2023}) near the propagation axis of the bunch.
% During the experiment with plasma, the screens upstream of the plasma entrance are extracted out of the beamline, and a laser beam dump is inserted to protect the screens downstream of the plasma exit.

\par
\begin{figure}[h!]
\centering
\includegraphics[scale=0.35]{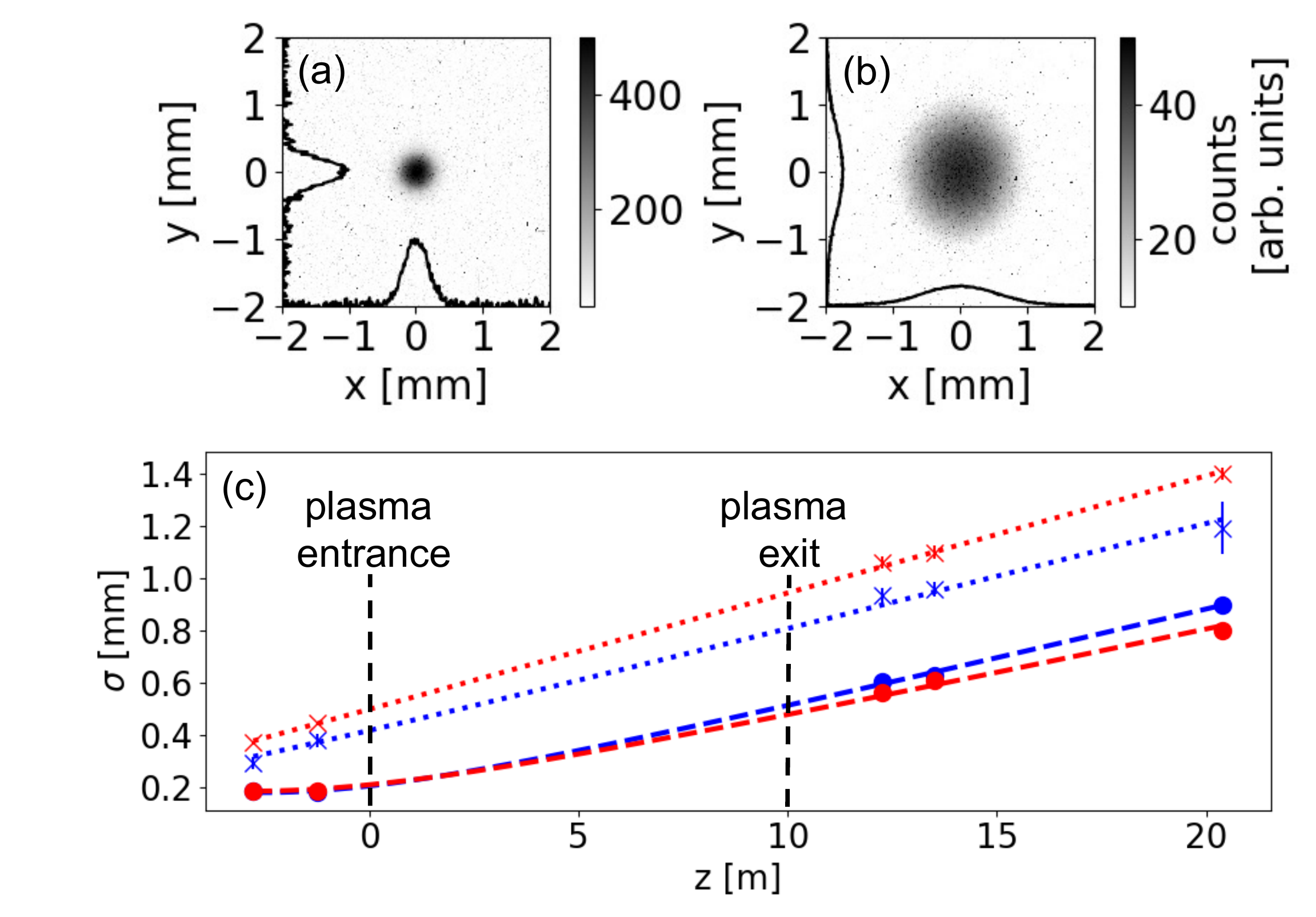}
\caption{
Transverse, time-integrated images ($x,y$) of (a) the narrow and (b) wide bunch, at $z= -1.5\,$m.
Black lines show the projections on the transverse planes.
(c): transverse size of the bunch along the beamline $\sigma_{x,y}$, obtained from Gaussian fits to the transverse projections of images at the various screens with no plasma.
Circles: narrow-bunch optics, crosses: wide-bunch optics; blue symbols: $x$-plane, red symbols: $y$-plane. 
Dashed and dotted lines: result of the fits with the envelope equation.
Plasma entrance at $z=0$, plasma exit at $z=10\,$m.
}
\label{fig:2}
\end{figure}

 Figures~\ref{fig:2}(a) and (b) show the time-integrated, transverse distribution of the bunch at the last screen before injection in plasma ($z= -1.5\,$m), for two sets of beam transport optics configurations, which we hereafter refer to as narrow and wide-bunch optics, respectively.
Fitting the transverse projections (black lines) with a Gaussian distribution, we calculate the transverse size of the beam in the two planes ($x,y$) as the rms $\sigma_{x,y}$.
The transverse size is also confirmed to be constant along the bunch using time-resolved images~\cite{LIVIO:PRL}.
% Figure~\ref{fig:2}(c) shows $\sigma_{x,y}$ measured at the various screens along the propagation axis for the two optical settings.
By fitting $\sigma_{x,y}$ measured without plasma at the various screens along the propagation axis $z$ (Fig.~\ref{fig:2}(c)) with the solution of the beam envelope equation in vacuum $\sigma_{x,y}(z) = \sigma^*_{x,y}\sqrt{1+ (z-z_{x,y}^*)^2(\frac{\epsilon_{x,y}/(\beta \gamma)}{\sigma_{x,y}^*2})^2}$, we obtain the transverse size at the waist $\sigma_{x,y}^*$, the waist position $z_{x,y}^*$ and the normalized transverse emittance $\epsilon_{x,y}$ ($\beta$ is the ratio between bunch longitudinal velocity and $c$, $\gamma$ is the Lorentz factor).
We then calculate the size of the beam at the plasma entrance $\sigma_{x,y}(z=0)$.
% Since the rms variation of transverse size at the screens is smaller than $5\,\%$ of the values, this is representative of the initial transverse size of the bunch when injecting in plasma and the measurement is not available.
For the narrow-bunch optics (circles in Fig.~\ref{fig:2}(c); blue: $x$-plane, red: $y$-plane; dashed lines: envelope equation fits), $\sigma_{x,y}^*=(0.19, 0.18)\,$mm, $\epsilon_{x,y}=(2.9,2.7)\,$mm-mrad, and the waist position is close to the plasma entrance: $z_{x,y}^*=(-2.6,-2.9)\,$m (smaller than the Twiss parameter $\beta_{x,y}^*=\sigma_{x,y}^{*2}\beta \gamma/\epsilon_{x,y}=(5.3,4.6)\,$m), and $\sigma_{x,y}(z=0)=(0.21,0.21)\,$mm.
For the wide-bunch optics (crosses and dotted lines in Fig.~\ref{fig:2}(c)), $z_{x,y}^*=(-9.4,-10.8)\,$m, with same emittance, and size at the waist as with the narrow bunch. 
This means that the waist is located about $2\beta_{x,y}^*$ upstream of the plasma entrance.
Thus, in this case the bunch enters the plasma diverging and with transverse size $\sigma_{x,y}=(0.42,0.50)\,$mm.
The uncertainties on the parameters of the fits are given in Ref.~\cite{uncertainties}.
As the difference in size between the two planes is smaller than the difference between the two optical settings, and it does not influence the results presented here, in the following we quote the average of the values in the two planes as $\sigma_{r0}$.
%%%%%%%%%
%%%%%%%%%%%%%%%%%%%%%%%%
\section{Experimental Results}
\begin{figure}[h!]
\centering
\includegraphics[scale=0.4]{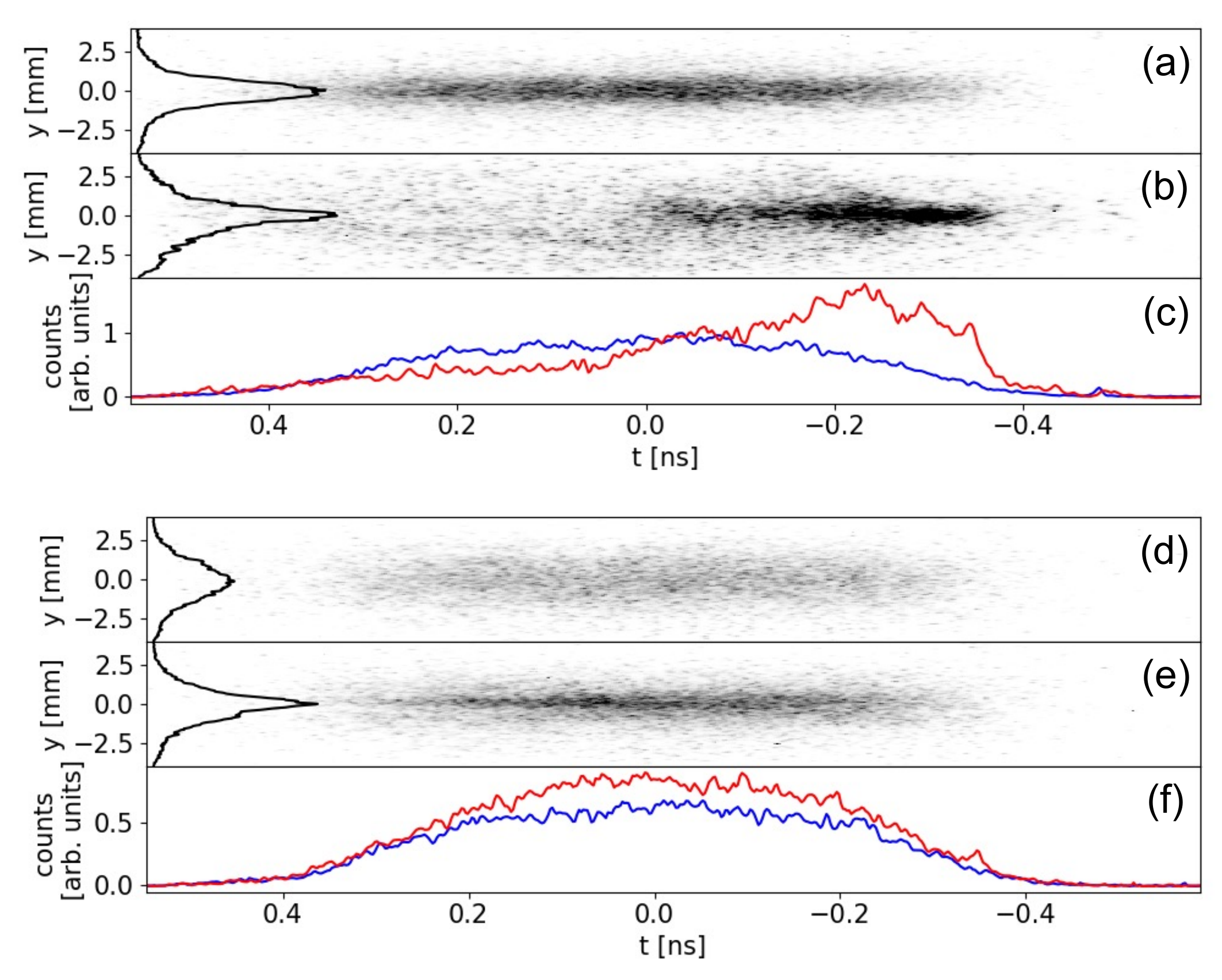}
\caption{
%%REWRITE
Single-event, time-resolved images of the $p^+$ bunch at the screen at $z=13.5\,$m.
(a) and (b): narrow bunch; (d) and (e): wide bunch. 
(a) and (d): after propagation without plasma; (b) and (e): after propagation with plasma. 
(c) and (f): time projections of narrow bunch and wide bunch, respectively, obtained by summing the counts along the $y$-spatial direction. Blue lines: without plasma, red lines: with plasma.
Black lines along the vertical axes: spatial projections.
The duration of the streak camera window is 1.1\,ns.
The bunch propagates from left to right; bunch center at $t=0$.
Same color scale for all images.
$n_{pe}=0.97\times 10^{14}\,$cm$^{-3}$.
}
\label{fig:3}
\end{figure}

% To study the growth of SMI along the bunch, we rely on time-resolved images of the $p^+$ bunch charge density distribution, obtained with the streak camera.
Figure~\ref{fig:3}(a) shows a single-event, time-resolved ns timescale image of the narrow $p^+$ bunch after propagation in vapor (no laser pulse, thus no plasma: bunch propagates as in vacuum), representing the incoming bunch as observed at the screen positioned 3.5\,m downstream of the vapor source exit.
The temporal (blue line in Fig.~\ref{fig:3}(c)) and spatial (black line in Fig.~\ref{fig:3}(a)) projections show that the bunch has a 2D-Gaussian charge density profile.
After propagation in plasma with $n_{pe}=0.97\times 10^{14}\,$cm$^{-3}$ ($\sigma_{r0}/(c/\omega_{pe})=0.39$ at the plasma entrance, see Fig.~\ref{fig:3}(b)), the spatial projection (black line) is no longer Gaussian, because of the occurrence of SMI: the microbunch train generates the bright core~\cite{KARL:PRL} and the defocused protons generate the surrounding halo~\cite{MARLENE:PRL}.

\par In this experiment, the laser pulse propagates $\sim 1\,$ns ahead of the $p^+$ bunch center ($\sim 4.7\,\sigma_z/c$), hence self-modulation occur as an instability, because the bunch density at the location of the ionization front is too low to seed~\cite{FABIAN:PRL}.
The time resolution of ns timescale images ($\sim 5\,$ps~\cite{Nechaeva_2023}) is not sufficient to resolve the microbunch structure at this plasma electron density (plasma electron period $T_{pe} = 2\pi/\omega_{pe}=11.3\,$ps).
Hence, the charge distribution appears continuous.
The transverse extent of the distribution along the bunch depends on the transverse momentum acquired by the defocused protons during the growth of the instability~\cite{MARLENE:PRAB,LIVIO:PRL}.
It is therefore a probe for the amplitude of the wakefields at the early stage of SMI.
The effect of the entrance slit of the streak camera is to decrease the intensity of the time-resolved images where the transverse extent of the distribution increases (the amount of charge per time slice remains constant)~\cite{Bachmann_2020}.
The time projection of the image with plasma (Fig.~\ref{fig:3}(c), red line) shows this effect by the higher count values at the front of the bunch ($-0.4\lesssim t\lesssim  -0.1\,$ns) than the projection of the image without plasma (blue line), because the bunch is focused by the adiabatic response of the plasma~\cite{LIVIO:PRL} and because of the formation of the microbunch train.
Conversely, the rapid decrease later along the bunch observed at the screen ($t>-0.2\,$ns) is due to defocusing from the occurrence of SMI.
%, causing the transverse extent of the bunch to increase. 
%%%%%%%%%
%%%%%%%%%%%%%%%%%%%%%%%%

\par In the case of the wide bunch (the charge and the other parameters of the bunch kept constant), the image without plasma (Fig.~\ref{fig:3}(d)) is less bright than for the narrow bunch (Fig.~\ref{fig:3}(a), the streak camera settings are the same for all images).
The spatial projection in the case with plasma (Fig.~\ref{fig:3}(e), black line; $\sigma_{r0}/(c/\omega_{pe})=0.85$) is peaked and no longer Gaussian.
% This may be due to SMI~\cite{MARLENE:PRL} and/or to a non-linear focusing force, such as the adiabatic response of the plasma~\cite{LIVIO:AAC22}. 
% Preliminary experimental results on the propagation of the $p^+$ bunch in low-density plasma ($n_{pe}\lesssim10^{12}\,$cm$^{-3}$), where the development of SMI is suppressed but the adiabatic response is the same, show that, with focusing only, the transverse distribution of the bunch remains essentially Gaussian~\cite{LIVIO:FOCUSING}.
We attribute the peaked transverse distribution of Fig.~\ref{fig:3}(e) to SMI, as confirmed later with ps timescale images (Fig.~\ref{fig:6}).
The time projection of the case with plasma (Fig.~\ref{fig:3}(f), red line) does not show a rapid decrease of counts along the bunch as in (c), indicating a weaker effect of SMI on the bunch and thus that the average amplitude of the transverse wakefields is much smaller than in the case of the narrow bunch.
In fact, in Fig.~\ref{fig:3}(e) there is no clear evidence of the surrounding halo distribution visible in Fig.~\ref{fig:3}(b).
This is expected because the smaller $n_{b0}$ and larger $\sigma_{r0}/(c/\omega_{pe})$ lead to smaller amplitude of the initial wakefields $W_{\perp 0}$ and smaller growth rate $\Gamma$~\cite{MARLENE:PRL, LIVIO:PRL}.

\par In the following Sections, we use a shorter streak camera time window (210\,ps), providing a better time resolution ($\sim2\,$ps) to evidence the microbunch train itself (if any) with narrow and wide bunch in low and high plasma electron densities.

\subsection{Narrow bunch, low plasma electron density}
\begin{figure}[h!]
\centering
\includegraphics[scale=0.42]{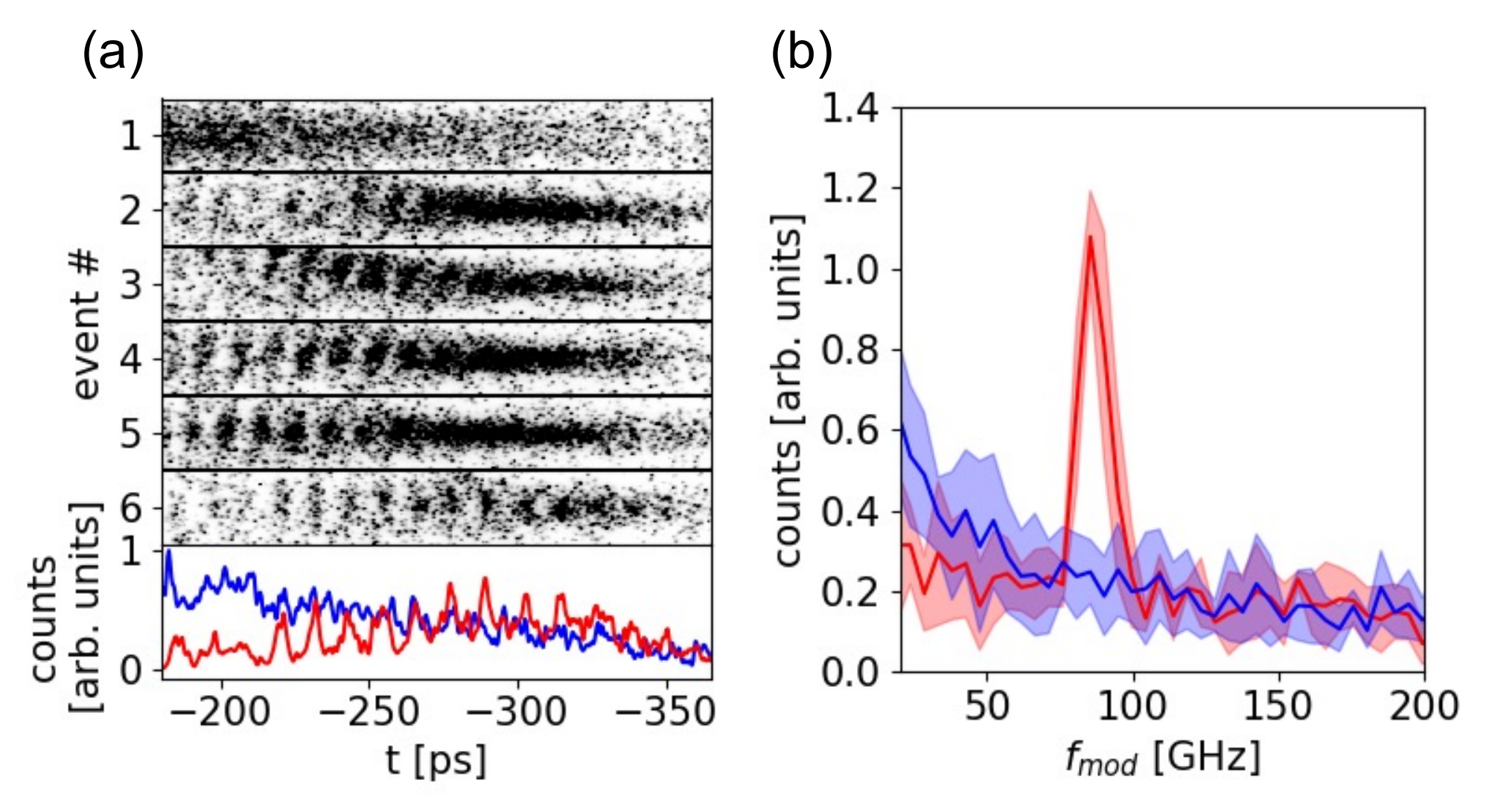}
\caption{
(a): six time-resolved, consecutive, single-event images of the narrow $p^+$ bunch.
Event~$\#1$: without plasma; events~$\#2-6$: with plasma. 
Blue line: on-axis profile of event~$\#1$; red line: on-axis profile of event~$\#6$.
Counts normalized with respect to the maximum of the projection of the image without plasma.
The duration of the streak camera window is 210\,ps.
The width of each image is $\Delta y=\pm 1.25\,$mm centered on the beam propagation axis.
(b): average power spectra obtained from the DFT of the on-axis profiles of single-event images. 
Red line: average of the power spectra of five consecutive events with plasma (events~$\#2-6$ from (a)); blue line: average of the power spectra of five consecutive events without plasma (including event~$\#1$).
The shaded areas show the extent of the rms variations.
Same color scale for all images.
$n_{pe}=0.97\times 10^{14}\,$cm$^{-3}$, $f_{pe}=88.3\,$GHz.
}
\label{fig:4}
\end{figure}

% To determine whether SMI occurs with the wide bunch, we use a shorter streak camera time window (210\,ps), providing a better time resolution ($\sim2\,$ps) to evidence the microbunch train itself (if any).
We observe the charge density distribution at the front of the bunch to investigate the position along the bunch where the modulation becomes detectable.
% which we demonstrate depends on $W_{\perp 0}$.
Figure~\ref{fig:4}(a) shows six consecutive, single-event, time-resolved images of the narrow $p^+$ bunch: one after propagation without plasma ($\#1$) and five after propagation with plasma ($\#2-6$), between 180 and 360\,ps ahead of the bunch center.
The images are aligned in time with sub-ps precision using an optical timing fiducial~\cite{FABIAN:MARKER}.
The on-axis profile of the event without plasma (blue line) shows the increase of the intensity along the Gaussian distribution of the bunch and no periodic oscillation.
% The rms jitter of the time of arrival of the $p^+$ bunch is $\sim 5\,$ps$\ll \sigma_z/c$, therefore negligible.
On the contrary, the microbunch structure is clearly visible on all events with plasma, and the corresponding on-axis profiles (e.g., red line, profile of event $\#6$ on Fig.~\ref{fig:4}(a)) show a periodic modulation.
% with maximum depth, defined as $M=(I_{max}-I_{min})/(I_{max}+I_{min})$ ($I_{max}$ and $I_{min}$ are the values of local maxima and minima, respectively), larger than $60\,\%$, for example at $t\sim -220\,$ps.
Images show that the timing of the modulation is not reproducible from event to event because the instability is not seeded~\cite{FABIAN:PRL}.

\par To measure the frequency of the modulation $f_{mod}$, we perform a discrete Fourier transform (DFT) of the on-axis profiles of the single-event images such as those of Fig.~\ref{fig:4}(a).
Figure~\ref{fig:4}(b) shows the average power spectra for five events with plasma (red line) and five events without plasma (blue line), together with their rms variation as the corresponding shaded areas.
The spectrum of the events with plasma is clearly peaked at $f_{mod}=85\pm5\,$GHz (the uncertainty is estimated as the width of a frequency bin), which is consistent with the expected plasma electron frequency $f_{pe}=\omega_{pe}/2\pi=88.3\,$GHz.
The signal-to-noise ratio, defined as the ratio between the peak value at $f_{mod}$ and the corresponding value for the spectrum without plasma, plus its rms, is close to three.
This indicates that the modulation depth of the profiles due to SMI is much larger than that due to the random variation in the distribution of the signal without plasma, as visible from the time-resolved images and profiles.
% Hence, at this location along the bunch, after $10\,$m of propagation in plasma, the exponential term $exp(\Gamma z)$ dominates on the amplitude of the wakefields and the microbunches are fully formed.
Hence, at this location along the bunch, after $10\,$m of propagation in plasma, the microbunches are fully formed, indicating that %
% exponential growth of 
the instability has taken place.
%%%%%%%%%
%%%%%%%%%%%%%%%%%%%%%%%%

\subsection{Wide bunch, low plasma electron density}

\begin{figure}[h!]
\centering
\includegraphics[scale=0.42]{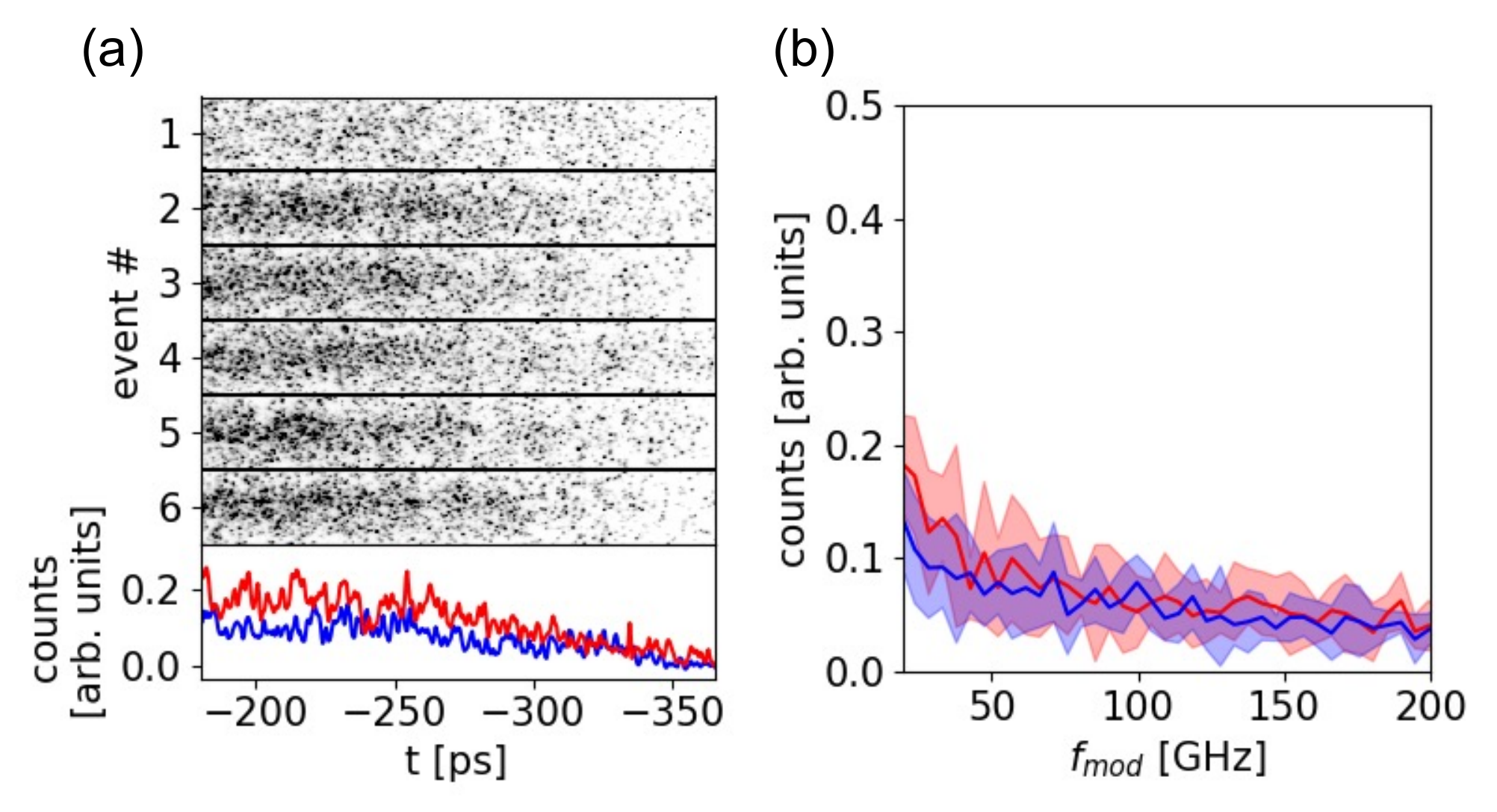}
\caption{
Same as Fig.~\ref{fig:4}, for the wide $p^+$ bunch.
}
\label{fig:5}
\end{figure}

In the case of the wide bunch (Fig.~\ref{fig:5}(a)), and over the same time range as for the narrow bunch (Fig.~\ref{fig:4}(a)), the microbunches are not distinguishable on time-resolved images.
% Images with plasma ($\#$2-6) have a brighter signal than the one without plasma ($\#$1) because of the effects of adiabatic focusing and of the slit, which is also visible from the on-axis profiles: the signal is larger with (red line) than without plasma (blue line).
The absence of the occurrence of SMI is also confirmed by the DFT analysis, that shows very similar power spectra with and without plasma (see Fig.~\ref{fig:5}(b)), i.e., without a detectable peak at the expected $f_{mod}\sim f_{pe}$.
This indicates that, if present, the periodic modulation due to SMI is not deeper than the variation due to noise in the distribution of the images obtained without plasma.

\par We estimate the threshold for detection of modulation in the power spectrum by calculating the amplitude of a sinusoidal modulation on a smooth ideal Gaussian distribution whose power spectrum near $f_{pe}$ has amplitude equal to twice the rms variation of the spectrum of the events without plasma.
% \par To calculate the minimum detectable modulation depth and estimate an upper limit to the effect of SMI, we apply the same DFT procedure on a Gaussian profile (representing a perfectly smooth Gaussian bunch in the same time range as in the time-resolved images), on which we add a sinusoidal component with frequency $f_{pe}$ (representing the modulation). 
% We define as detection threshold the amplitude of this component for which the peak of the power spectrum is equal to that of the average spectrum of the events without plasma.
In the case of Fig.~\ref{fig:5}(b), the threshold, and therefore the lower limit for detection of the modulation depth due to SMI, is $\sim20\,\%$.
%(relative to the amplitude of the Gaussian distribution).
For comparison, according to the same simple model, the peak of the power spectrum of the narrow bunch (Fig.~\ref{fig:4}(b)) corresponds to an amplitude of the sinusoidal component $>40\,\%$, with detection threshold $\sim15\,\%$.
%%%%%%%%%
%%%%%%%%%%%%%%%%%%%%%%%%
\par
\begin{figure}[h!]
\centering
\includegraphics[scale=0.42]{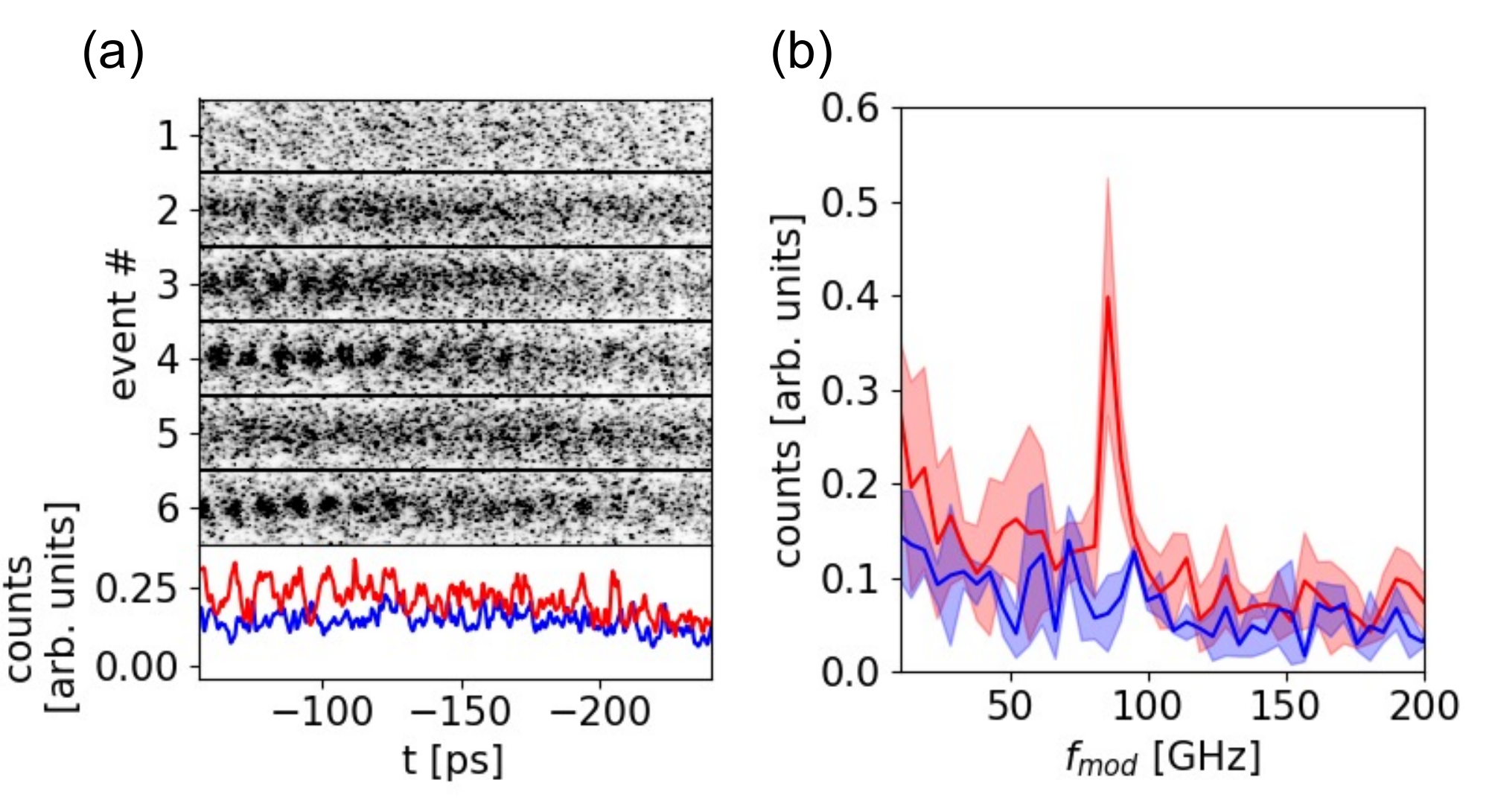}
\caption{
Same as Fig.~\ref{fig:5}, for a later time range along the wide $p^+$ bunch.
}
\label{fig:6}
\end{figure}

 Figure~\ref{fig:6}(a) shows that, with the same wide bunch, microbunches do become visible later along the bunch ($-240\leq t\leq-60\,$ps, rather than $-360\leq t\leq-180\,$ps as in Fig.~\ref{fig:4}(a)).
The average power spectrum (red line, Fig.~\ref{fig:6}(b)) shows a clear peak above the noise level at $f_{mod}\sim f_{pe}$, indicating a modulation depth around four times larger than the minimum detectable value ($\sim 15\,\%$, estimated as for Fig.~\ref{fig:5}).
% , in agreement with the maximum modulation depth $M\sim 40\,\%$ measurable on the on-axis projection of a single-event image with plasma from Fig.~\ref{fig:6}(a) (red line, event $\#6$, at $t\sim -50\,$ps).

\par This confirms that SMI also takes place with the wide bunch (as already suggested by the spatial projection of Fig.~\ref{fig:3}(e)), but it is observed only later along the bunch than with the narrow bunch.
This is because the amplitude of the initial wakefields $W_{\perp0}$ is smaller, with the smaller $n_{b0}$ and larger $\sigma_{r0}/(c/\omega_{pe})$.
% Hence, it takes longer for the transverse wakefields to overcome the divergence of the bunch and to generate the initial modulation needed for the instability to start growing.
This also indicates a longer saturation length of the instability as well as a lower amplitude of the wakefields driven by the bunch, at any given location along the plasma and bunch.
% \textcolor{red}{The fact that the beam enters the plasma diverging also affects the development of SMI, with the same general effect as a larger emittance~\cite{GORN:DEFOCUSING}, as it takes a longer distance for the focusing force to generate the initial modulation. 
% In the following we show that the same effect is obtained when increasing $n_{pe}$, keeping the parameters of the bunch unchanged.}

%%%%%%%%%
%%%%%%%%%%%%%%%%%%%%%%%%

\subsection{High plasma electron density}

\begin{figure}[h!]
\centering
\includegraphics[scale=0.4]{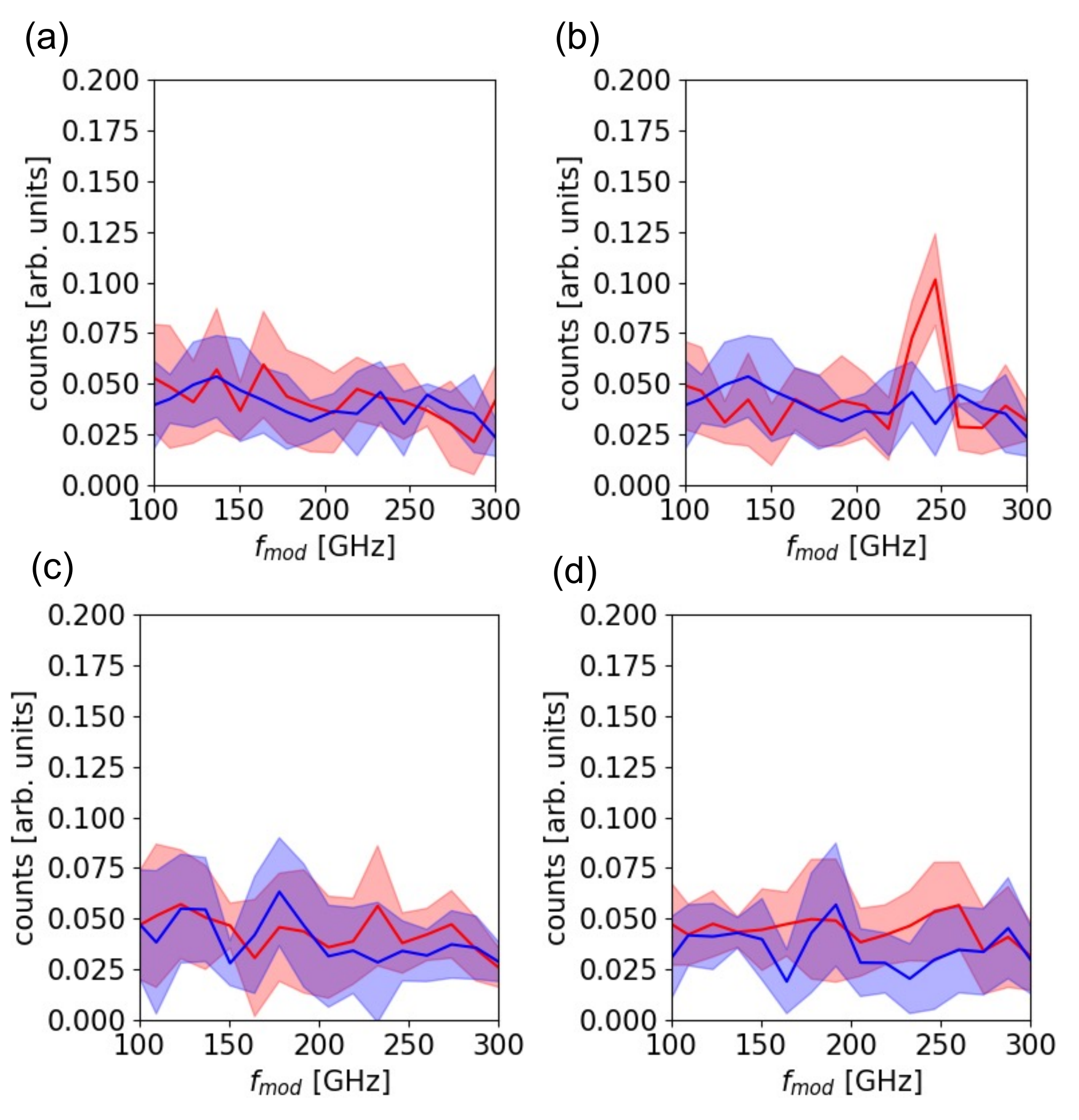}
\caption{
Average power spectra from the DFT analysis on the on-axis profiles of single-events images.
Red lines: average of five consecutive events with plasma; blue lines: average of five consecutive events without plasma.
The shaded areas show the extent of the corresponding rms variations.
The duration of the streak camera window is 73\,ps for all cases.
(a): narrow bunch, measuring between $t=-300$ and $-370\,$ps (early).
(b): narrow bunch, measuring between $t=-150$ and $-220\,$ps (late).
(c) wide bunch, measuring between $t=-150$ and $-220\,$ps.
(d) wide bunch, RIF at the bunch center, measuring between $t=+170$ and $+240\,$ps (i.e., behind the center of the bunch).
$n_{pe}=7.3\times 10^{14}\,$cm$^{-3}$, $f_{pe}=242.3\,$GHz.
}
\label{fig:7}
\end{figure}

The plasma skin depth shortens when increasing $n_{pe}$.
Thus, for a fixed transverse size of the drive bunch, the ratio $\sigma_{r0}/(c/\omega_{pe})$ increases, less bunch charge is contained within $c/\omega_{pe}$, and the amplitude of the initial wakefields $W_{\perp 0}$ decreases.
We acquire time-resolved images of the bunch with $n_{pe}=7.3\times 10^{14}\,$cm$^{-3}$ in a 73 ps time window, with sufficient time resolution ($\sim 1\,$ps) to detect modulation at the expected $f_{pe}=242.3\,$GHz ($T_{pe}=4.13\,$ps), at the expense of a lower signal-to-noise ratio because of lower counts per pixel.
Single images do not show evidence of SMI due to the limited time resolution. 
We therefore only use the more sensitive DFT method to assess the occurrence of SMI~\cite{KARL:STREAK}.
%In the following we show only results from DFT analysis of time profiles.
Figure~\ref{fig:7}(a) shows that the average power spectrum obtained from time-resolved images of the narrow bunch in plasma ($\sigma_{r0}/(c/\omega_{pe})=1.1$, red line) at $-370\leq t\leq-300\,$ps (i.e., shorter than, but within the same time range as Fig.~\ref{fig:4}) does not have a peak with amplitude significantly larger than the level of the average spectrum of the events without plasma (blue).
This indicates a possible modulation depth no larger than the initial rms variation of the bunch density distribution ($\sim 20\,\%$).

\par At $-220\leq t \leq -150\,$ps (i.e., later along the bunch, Fig.~\ref{fig:7}(b)), the average power spectrum of the images with plasma is clearly peaked at $f_{mod}=246\pm 14\,$GHz\,$\sim f_{pe}$.
The uncertainty is larger than in the previous cases because the observation window is shorter.
%This indicates that the effect of lower $W_{\perp 0}$ due to the larger $\sigma_{r0}/(c/\omega_{pe})$ dominates the development of SMI, despite the larger $\Gamma$ at this higher $n_{pe}$~\cite{KUMAR:GROWTH,PUKHOV:GROWTH,SCH:GROWTH}.
Observing a peak in the power spectrum confirms that the lower $W_{\perp 0}$ causes a later development of SMI along the bunch.
% despite the larger $\Gamma$ at this higher $n_{pe}$~\cite{KUMAR:GROWTH,PUKHOV:GROWTH,SCH:GROWTH}.

\par With the wide bunch (Fig.~\ref{fig:7}(c)), there is no visible peak at $f_{pe}$ with amplitude higher than the threshold ($\sim 20\,\%$) in the same late time window as for (b), indicating again that lower $n_{b0}$  and larger $\sigma_{r0}/(c/\omega_{pe})$ delay the development of the microbunch train along the bunch.
To further investigate whether the wide bunch undergoes SMI at all in high-density plasma, we also measure the charge density distribution when the ionization front propagates at the center of the bunch, i.e., we impose the maximum possible amplitude of the initial wakefields (RIF seeding).
Figure~\ref{fig:7}(d) shows that no peak at $f_{pe}$ in the average power spectrum of the events with plasma is distinguishable from the power spectrum of the events without plasma ($+170\leq t \leq +240\,$ps).
We therefore conclude that at this high density the wide bunch does not self-modulate at all over 10\,m, even when SMI is strongly seeded.
In this case $\sigma_{r0}/(c/\omega_{pe})=2.4>1$, which means that the plasma return current can flow through the bunch and current filamentation instability~\cite{CFI:PhysRevLett.31.1390,CFI:PhysRevLett.109.185007} could develop.
Further experiments to study this phenomenon and its effects are underway~\cite{PATRIC:IPAC23}.
%%%%%%%%%
%%%%%%%%%%%%%%%%%%%%%%%%
\subsection{Summary of the Experimental Results}

% The experimental results we presented show the competing effect between emittance-related bunch divergence and focusing wakefields on the development of SMI.
% They show that SMI does not occur for example with the largest initial transverse size and plasma electron density available in our experiments.
% However, as noted, noise in the time-resolved images imposes a detection limitation. 
% Thus, even though SMI is not detected, it may be present at a level lower than the detection threshold. 
% Also, we observed hints of CFI. 
% Therefore we cannot exclude that these  instabilities appear in the second plasma. 
% 

% The experimental results we present here show the competing effect between bunch divergence due to emittance and focusing wakefields on the development of SMI.
% They suggest that SMI develops only after a given point along the bunch. 

% They show that SMI does not occur for example with the largest initial transverse size and plasma electron density available in our experiments.

\par Previous measurements~\cite{FABIAN:PRL} indicate that SMI ``starts" at a location along a Gaussian bunch where a perturbation of, or imperfection in the bunch density (on a spatial scale $\ll c/\omega_{pe}$), limited by the local bunch density, is sufficient to initiate the process. % 
This occurs where the amplitude of the wakefields that the perturbation drives is sufficient for the resulting transverse force to overcome the effect of the divergence of the slice with finite emittance. %
While each slice of the bunch, with the same radius and emittance, diverges at the same rate at the entrance of the plasma, the amplitude of wakefields that can be driven increases from the front of the bunch till its peak. %
One can estimate a first position along the bunch where SMI may occur by using the amplitude of wakefields driven if there were a sharp step in the bunch density at the location of the slice. %
In this case, the amplitude $W_{\perp0}$ can be calculated as was done in Ref.~\cite{FABIAN:PRL} using linear wakefield theory, since $n_{b0}\ll n_{pe}$. %
This is equivalent to using an envelope equation for each slice at the entrance of the plasma: %
\begin{equation}
\label{eq:matching}
\sigma_r''=\left(\frac{\epsilon_g^2}{\sigma_{r0}^3}-\frac{eW_{\perp0}}{\gamma m_ec^2}\right),
\end{equation}
where $W_{\perp0}(\xi)=2\frac{en_{b0}}{\epsilon_0k_{pe}^2}\exp{(-\xi/2\sigma_{z}^2)} dR/dr |_{r=\sigma_{r0}}$.%
The $R(r)$ coefficient is a function of the transverse bunch profile and describes the radial dependence of the wakefields~\cite{JONES:LIN_THEORY}.
The position along the bunch where SMI can first develop from, can be calculated by setting the RHS of Eq.~\ref{eq:matching} to zero, to find the equivalent of a matching condition.
%(bunch not at a waist at the plasma entrance). 
For each slice ahead of it, the effect of divergence dominates (i.e., RHS$>0$) and SMI cannot develop. %
For each slice behind it, RHS$<$0 and SMI can develop. %
In particular, in a case of a bunch for which the ``matched" condition cannot be met before the peak of the bunch, SMI can never develop. %
This is because, unlike in the classical case of focusing in a pure ion column where the force increases (linearly) with radius~\cite{BLOWOUT:ROSENZ, PATRIC:MATCHING}, the focusing force generated by diverging slices decreases with propagation distance ($W_{\perp}(z)\propto n_{b0}(z)\propto1/\sigma_r^2(z)$), ensuring that past the plasma entrance SMI never develops ahead of the "matched" slice.
This is confirmed by the cases of Fig.~\ref{fig:7}(c), for which SMI does not develop, and Fig.~\ref{fig:7}(d), for which it does not even with the strongest possible $W_{\perp0}$ (note that in this case the bunch is not at a waist at the plasma entrance).

\par
\begin{figure}[h!]
\centering
\includegraphics[scale=0.4]{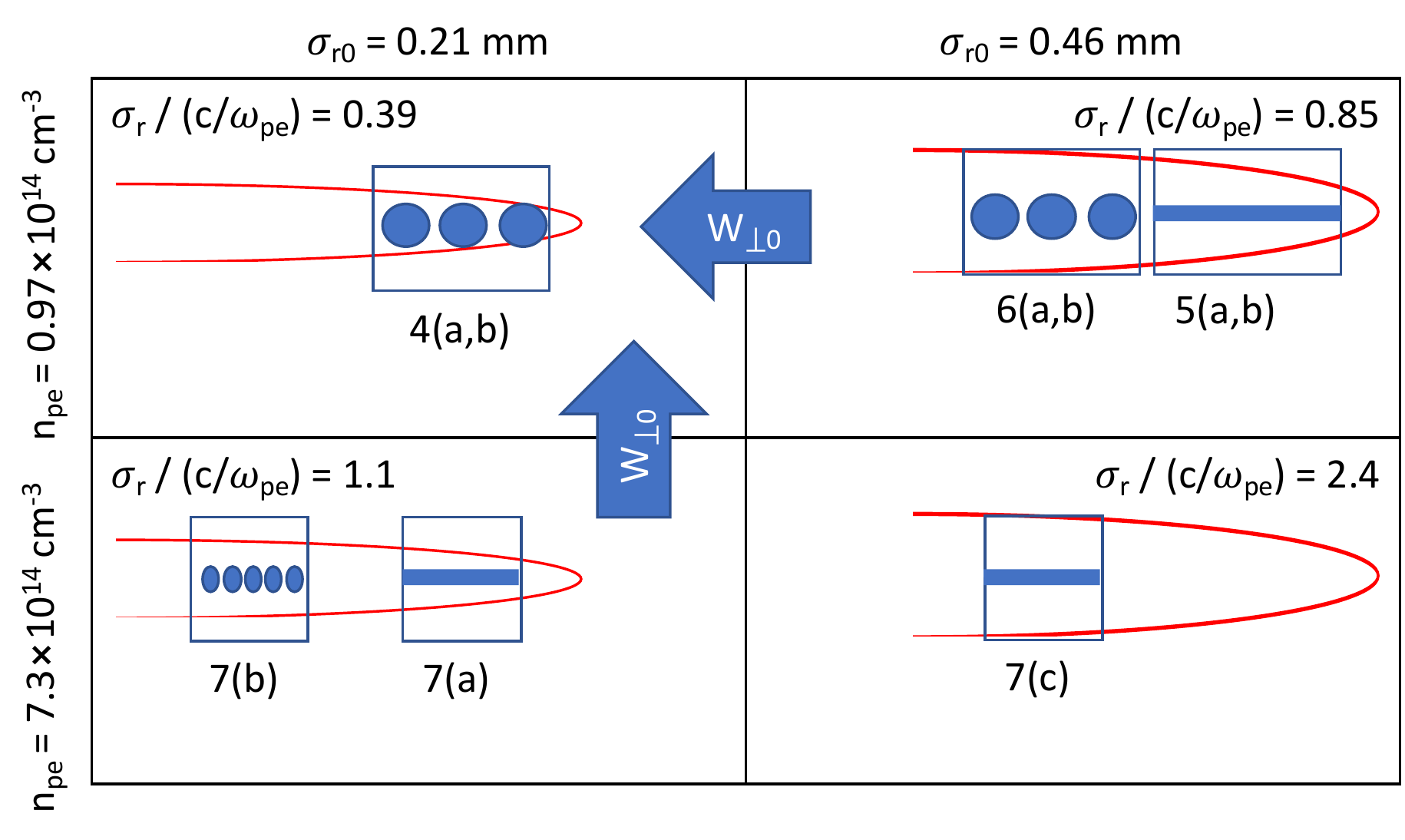}
\caption{
Schematic summary of the experimental results. 
The red half ellipses represent the $p^+$ bunch traveling from left to right, the blue rectangles represent the streak camera windows where measurements were performed. 
Blue circles: SMI observed in the window; blue lines: SMI not observed.
The blue arrows indicate how $W_{\perp0}$ varies with $\sigma_{r0}$ (horizontal direction) and with $n_{pe}$ (vertical direction).
}
\label{fig:8}
\end{figure}

 We summarize the results in Fig.~\ref{fig:8} by schematically displaying what was observed in each time window (blue circles: SMI, continuous line: no SMI) from time-resolved images and from their DFT power spectrum (i.e., detecting or not a peak corresponding to the modulation).
% SMI occurs later, or not at all, with the wider bunch (horizontal direction in the scheme), simply because a wider bunch drives initial wakefields with lower amplitude.
% The same effect occurs with higher plasma electron density (vertical direction).
% In the first case, the growth rate is also decreased, while in the second case it is actually increased. 
The results indicate that the earliest time along the bunch where SMI is observed depends on the amplitude of the initial wakefields $W_{\perp0}$, in agreement with what is suggested by the simple model we presented. 
When increasing the transverse size of the bunch (horizontal direction in the schematic) or when increasing the plasma electron density (vertical direction), the amplitude of the initial wakefields decreases.
Regardless of the growth rate, SMI can only occur where and when the initial wakefields overcome the divergence of the bunch.
However, as noted, noise in the time-resolved images imposes a detection limitation. 
Thus, even though SMI is not detected, it may be present at a level lower than the detection threshold. 
%%%%%%%%%
%%%%%%%%%%%%%%%%%%%%%%%%

\section{Impact on accelerator design}
\label{sec:run2c}
To effectively employ the $p^+$-driven scheme for accelerator applications, the timing and amplitude of the wakefields must be reproducible from event to event, because the witness bunch must be deterministically injected in the accelerating and focusing phase of the wakefields with sub-ps accuracy for every event. 
Reproducibility is obtained by seeding the instability~\cite{FANG:PhysRevLett.112.045001,FABIAN:PRL,LIVIO:PRL}, i.e., driving initial wakefields with sufficient amplitude for SMI to grow from.
% We showed in previous experiments that SMI can be seeded by a relativistic ionization front (RIF) copropagating within the $p^+$ bunch~\cite{FABIAN:PRL}, or by the wakefields driven by a preceding, short electron bunch~\cite{LIVIO:PRL}.
% , the seed wakefields are provided by the fast onset of the beam-plasma interaction at the location of the RIF.
In case of RIF seeding, a transition from the instability to the seeded regime was observed when the local bunch density at the RIF location overcomes a threshold value~\cite{FABIAN:PRL}.
For the narrow bunch, we measured the transition to occur between 400 and $300\,$ps ahead of the bunch center (not shown here). 

\par In one of the possible designs of the accelerator employing the self-modulated $p^+$ bunch as the driver of plasma wakefields~\cite{EDDA:IPAC,PATRIC:EAAC}, SMI is made reproducible by seeding in a first plasma with a RIF.
In this case, the part of the bunch ahead of RIF keeps propagating in vapor, diverging as in vacuum.
Then, the bunch enters a 10-m-long preformed plasma (the accelerator), after a gap region where the on-axis injection of the witness electron bunch takes place~\cite{LIVIO:EAAC}.

\par With the experiments presented in this paper, we replicate with the wide bunch and a single plasma the conditions of the bunch front entering the second plasma, with large initial transverse size and similar divergence. 
The results show that, after propagation through 10\,m of plasma, the radial and longitudinal intensity modulation ahead of the transition point for RIF seeding~\cite{FABIAN:PRL} does not reach a depth distinguishable from that of the incoming bunch distribution (see Figs.~\ref{fig:5} and~\ref{fig:7}(c)), after propagation in the 10-m-long plasma that corresponds to the second plasma of the accelerator setup.
% , i.e. SMI does not develop.
This suggests that acceleration driven only by the self-modulated back part of the bunch can proceed without interference caused by SMI of the front of the bunch in the second plasma~\cite{Baistrukov_2022}. 
This is a necessary condition for producing a high-quality accelerated bunch with this scheme.
% This means that, at least over 10\,m, the front of the bunch traveling in the second plasma will not drive wakefields with amplitude high enough to disrupt the structure of the (seeded) self-modulated microbunch structure.
The acceleration experiments with the two plasma sections are anyway needed to prove that neither SMI nor CFI occur in the second plasma.

% \par In case of $n_{pe}=1$, we observe SMI to develop later along the bunch.
% By scaling $\Gamma\propto (n_{b0} z^2)^{1/3}$, we can expect the microbunch train at any given location along the bunch front to reach the same the same depth as the narrow bunch not before 20\,m of propagation in plasma.
% This may be even longer for $n_{pe}=7$ (design value for AWAKE), since we do not observe SMI after $z=10\,$m.

% Thus, the front of the bunch entering the second plasma section should not cause a deteriorating effect on the self-modulated (RIF seeded) part, and on the acceleration process.
\section{Conclusions}

We showed with experimental results that the condition for SMI to develop along a long $p^+$ bunch depends on the amplitude of the initial wakefields that the bunch can drive. 
This amplitude decreases when increasing the ratio $\sigma_{r0}/(c/\omega_{pe})\propto\sigma_{r0}\sqrt{n_{pe}}$.
We observed that SMI appears later along the bunch when increasing the initial transverse size of the bunch or the plasma electron density, and does not occur (within the detection threshold) when increasing both. 
% We introduced a simple model showing that SMI would not develop after any plasma length, at the locations along the bunch where the effect of the initial transverse wakefields does not compensate for the divergence of the beam.
Measurements also confirmed that, when increasing the initial transverse size $\sigma_{r0}$ (thus, decreasing $n_{b0}$), while keeping the other parameters constant, the growth rate $\Gamma$ of the self-modulation instability (when this occurs) decreases.
% We also show that, when increasing the ratio $\sigma_r/(c/\omega_{pe})$ by increasing $\sigma_r$ or $n_{pe}$, SMI  becomes observable later along the bunch, over a fixed plasma length, because the amplitude of the initial wakefields $W_{\perp 0}$ decreases: it takes longer for the initial modulation to form and for the instability to start growing.
\par We discussed the impact of these results on the design of a $p^+$-driven plasma wakefield accelerator based on the self-modulation instability seeded by a copropagating relativistic ionization front. 
We showed that, for a diverging bunch with large transverse size (comparable to that of the bunch front left unmodulated from a first plasma where seeding occurs), the modulation at the bunch front does not reach a detectable depth ahead of the transition point for RIF seeding. 
% This means that RIF seeding may be suitable for acceleration applications.
% Further experiments with a second plasma and CFI studies are needed to confirm

% When increasing $n_{pe}$, $\Gamma$ increases~\cite{KUMAR:GROWTH,PUKHOV:GROWTH,SCH:GROWTH}, but $W_{\perp0}$ decreases because the ratio $\sigma_{x,y}/(c/\omega_{pe})$ is larger. 

% This is due to the growth rate of the instability $\Gamma$ and the initial transverse wakefields $W_{\perp 0}$ being smaller for smaller $n_{b0}$.
% Moreover, the development of SMI is further dwarfed when $\sigma_{x,y}/(c\omega_{pe})>1$ because only a fraction of the bunch contributes to drive wakefields on axis. 
% This physics result has an important impact on the future design of the AWAKE experiment~\cite{PATRIC:EAAC,EDDA:IPAC}, because they show that the unmodulated front of the bunch leaving the first plasma would not self-modulate in the second plasma and would not affect the self-modulated (RIF seeded) following part of the bunch, driving large-amplitude wakefields for particle acceleration.

\begin{acknowledgments}
This work was supported in parts by STFC (AWAKE-UK, Cockcroft Institute core, John Adams Institute core, and UCL consolidated grants), United Kingdom,
the National Research Foundation of Korea (Nos.\ NRF-2016R1A5A1013277 and NRF-2020R1A2C1010835).
M. Wing acknowledges the support of DESY, Hamburg.
Support of the National Office for Research, Development and Innovation (NKFIH) under contract numbers 2019-2.1.6-NEMZ\_KI-2019-00004, and the support of the Wigner Datacenter Cloud facility through the Awakelaser project is acknowledged.
TRIUMF contribution is supported by NSERC of Canada.
The AWAKE collaboration acknowledge the SPS team for their excellent proton delivery.
\end{acknowledgments}

\section*{Data Availability Statement}
The data that support the findings of this study are available from the corresponding author upon reasonable request.

\bibliography{aipsamp}% Produces the bibliography via BibTeX.

%merlin.mbs aipnum4-1.bst 2010-07-25 4.21a (PWD, AO, DPC) hacked
%Control: key (0)
%Control: author (8) initials jnrlst
%Control: editor formatted (1) identically to author
%Control: production of article title (0) allowed
%Control: page (1) range
%Control: year (1) truncated
%Control: production of eprint (0) enabled
\begin{thebibliography}{31}%
\makeatletter
\providecommand \@ifxundefined [1]{%
 \@ifx{#1\undefined}
}%
\providecommand \@ifnum [1]{%
 \ifnum #1\expandafter \@firstoftwo
 \else \expandafter \@secondoftwo
 \fi
}%
\providecommand \@ifx [1]{%
 \ifx #1\expandafter \@firstoftwo
 \else \expandafter \@secondoftwo
 \fi
}%
\providecommand \natexlab [1]{#1}%
\providecommand \enquote  [1]{``#1''}%
\providecommand \bibnamefont  [1]{#1}%
\providecommand \bibfnamefont [1]{#1}%
\providecommand \citenamefont [1]{#1}%
\providecommand \href@noop [0]{\@secondoftwo}%
\providecommand \href [0]{\begingroup \@sanitize@url \@href}%
\providecommand \@href[1]{\@@startlink{#1}\@@href}%
\providecommand \@@href[1]{\endgroup#1\@@endlink}%
\providecommand \@sanitize@url [0]{\catcode `\\12\catcode `\$12\catcode
  `\&12\catcode `\#12\catcode `\^12\catcode `\_12\catcode `\%12\relax}%
\providecommand \@@startlink[1]{}%
\providecommand \@@endlink[0]{}%
\providecommand \url  [0]{\begingroup\@sanitize@url \@url }%
\providecommand \@url [1]{\endgroup\@href {#1}{\urlprefix }}%
\providecommand \urlprefix  [0]{URL }%
\providecommand \Eprint [0]{\href }%
\providecommand \doibase [0]{http://dx.doi.org/}%
\providecommand \selectlanguage [0]{\@gobble}%
\providecommand \bibinfo  [0]{\@secondoftwo}%
\providecommand \bibfield  [0]{\@secondoftwo}%
\providecommand \translation [1]{[#1]}%
\providecommand \BibitemOpen [0]{}%
\providecommand \bibitemStop [0]{}%
\providecommand \bibitemNoStop [0]{.\EOS\space}%
\providecommand \EOS [0]{\spacefactor3000\relax}%
\providecommand \BibitemShut  [1]{\csname bibitem#1\endcsname}%
\let\auto@bib@innerbib\@empty
%</preamble>
\bibitem [{\citenamefont {Kumar}, \citenamefont {Pukhov},\ and\ \citenamefont
  {Lotov}(2010)}]{KUMAR:GROWTH}%
  \BibitemOpen
  \bibfield  {author} {\bibinfo {author} {\bibfnamefont {N.}~\bibnamefont
  {Kumar}}, \bibinfo {author} {\bibfnamefont {A.}~\bibnamefont {Pukhov}}, \
  and\ \bibinfo {author} {\bibfnamefont {K.}~\bibnamefont {Lotov}},\ }\bibfield
   {title} {\enquote {\bibinfo {title} {Self-modulation instability of a long
  proton bunch in plasmas},}\ }\href {\doibase 10.1103/PhysRevLett.104.255003}
  {\bibfield  {journal} {\bibinfo  {journal} {Physical Review Letters}\
  }\textbf {\bibinfo {volume} {104}},\ \bibinfo {pages} {255003} (\bibinfo
  {year} {2010})}\BibitemShut {NoStop}%
\bibitem [{\citenamefont {{AWAKE Collaboration}}(2019)}]{KARL:PRL}%
  \BibitemOpen
  \bibfield  {author} {\bibinfo {author} {\bibnamefont {{AWAKE
  Collaboration}}},\ }\bibfield  {title} {\enquote {\bibinfo {title}
  {Experimental observation of proton bunch modulation in a plasma at varying
  plasma densities},}\ }\href {\doibase 10.1103/PhysRevLett.122.054802}
  {\bibfield  {journal} {\bibinfo  {journal} {Physical Review Letters}\
  }\textbf {\bibinfo {volume} {122}},\ \bibinfo {pages} {054802} (\bibinfo
  {year} {2019})}\BibitemShut {NoStop}%
\bibitem [{\citenamefont {{{M. Turner et al.}}}(2019)}]{MARLENE:PRL}%
  \BibitemOpen
  \bibfield  {author} {\bibinfo {author} {\bibnamefont {{{M. Turner et al.}}}}
  (\bibinfo {collaboration} {AWAKE Collaboration}),\ }\bibfield  {title}
  {\enquote {\bibinfo {title} {Experimental observation of plasma wakefield
  growth driven by the seeded self-modulation of a proton bunch},}\ }\href
  {\doibase 10.1103/PhysRevLett.122.054801} {\bibfield  {journal} {\bibinfo
  {journal} {Physical Review Letters}\ }\textbf {\bibinfo {volume} {122}},\
  \bibinfo {pages} {054801} (\bibinfo {year} {2019})}\BibitemShut {NoStop}%
\bibitem [{\citenamefont {{{F. Batsch, P. Muggli et al.}}}(2021)}]{FABIAN:PRL}%
  \BibitemOpen
  \bibfield  {author} {\bibinfo {author} {\bibnamefont {{{F. Batsch, P. Muggli
  et al.}}}} (\bibinfo {collaboration} {AWAKE Collaboration}),\ }\bibfield
  {title} {\enquote {\bibinfo {title} {Transition between instability and
  seeded self-modulation of a relativistic particle bunch in plasma},}\ }\href
  {\doibase 10.1103/PhysRevLett.126.164802} {\bibfield  {journal} {\bibinfo
  {journal} {Physical Review Letters}\ }\textbf {\bibinfo {volume} {126}},\
  \bibinfo {pages} {164802} (\bibinfo {year} {2021})}\BibitemShut {NoStop}%
\bibitem [{\citenamefont {Lotov}\ \emph {et~al.}(2013)\citenamefont {Lotov},
  \citenamefont {Lotova}, \citenamefont {Lotov}, \citenamefont {Upadhyay},
  \citenamefont {T\"uckmantel}, \citenamefont {Pukhov},\ and\ \citenamefont
  {Caldwell}}]{KOSTANTIN:NOISE}%
  \BibitemOpen
  \bibfield  {author} {\bibinfo {author} {\bibfnamefont {K.~V.}\ \bibnamefont
  {Lotov}}, \bibinfo {author} {\bibfnamefont {G.~Z.}\ \bibnamefont {Lotova}},
  \bibinfo {author} {\bibfnamefont {V.~I.}\ \bibnamefont {Lotov}}, \bibinfo
  {author} {\bibfnamefont {A.}~\bibnamefont {Upadhyay}}, \bibinfo {author}
  {\bibfnamefont {T.}~\bibnamefont {T\"uckmantel}}, \bibinfo {author}
  {\bibfnamefont {A.}~\bibnamefont {Pukhov}}, \ and\ \bibinfo {author}
  {\bibfnamefont {A.}~\bibnamefont {Caldwell}},\ }\bibfield  {title} {\enquote
  {\bibinfo {title} {Natural noise and external wakefield seeding in a
  proton-driven plasma accelerator},}\ }\href {\doibase
  10.1103/PhysRevSTAB.16.041301} {\bibfield  {journal} {\bibinfo  {journal}
  {Physical Review Special Topics Accelators and Beams}\ }\textbf {\bibinfo
  {volume} {16}},\ \bibinfo {pages} {041301} (\bibinfo {year}
  {2013})}\BibitemShut {NoStop}%
\bibitem [{\citenamefont {Chen}\ \emph {et~al.}(1985)\citenamefont {Chen},
  \citenamefont {Dawson}, \citenamefont {Huff},\ and\ \citenamefont
  {Katsouleas}}]{PWFA:CHEN}%
  \BibitemOpen
  \bibfield  {author} {\bibinfo {author} {\bibfnamefont {P.}~\bibnamefont
  {Chen}}, \bibinfo {author} {\bibfnamefont {J.~M.}\ \bibnamefont {Dawson}},
  \bibinfo {author} {\bibfnamefont {R.~W.}\ \bibnamefont {Huff}}, \ and\
  \bibinfo {author} {\bibfnamefont {T.}~\bibnamefont {Katsouleas}},\ }\bibfield
   {title} {\enquote {\bibinfo {title} {Acceleration of electrons by the
  interaction of a bunched electron beam with a plasma},}\ }\href {\doibase
  10.1103/PhysRevLett.54.693} {\bibfield  {journal} {\bibinfo  {journal}
  {Physical Review Letters}\ }\textbf {\bibinfo {volume} {54}},\ \bibinfo
  {pages} {693--696} (\bibinfo {year} {1985})}\BibitemShut {NoStop}%
\bibitem [{\citenamefont {Pukhov}\ \emph {et~al.}(2011)\citenamefont {Pukhov},
  \citenamefont {Kumar}, \citenamefont {T\"uckmantel}, \citenamefont
  {Upadhyay}, \citenamefont {Lotov}, \citenamefont {Muggli}, \citenamefont
  {Khudik}, \citenamefont {Siemon},\ and\ \citenamefont
  {Shvets}}]{PUKHOV:GROWTH}%
  \BibitemOpen
  \bibfield  {author} {\bibinfo {author} {\bibfnamefont {A.}~\bibnamefont
  {Pukhov}}, \bibinfo {author} {\bibfnamefont {N.}~\bibnamefont {Kumar}},
  \bibinfo {author} {\bibfnamefont {T.}~\bibnamefont {T\"uckmantel}}, \bibinfo
  {author} {\bibfnamefont {A.}~\bibnamefont {Upadhyay}}, \bibinfo {author}
  {\bibfnamefont {K.}~\bibnamefont {Lotov}}, \bibinfo {author} {\bibfnamefont
  {P.}~\bibnamefont {Muggli}}, \bibinfo {author} {\bibfnamefont
  {V.}~\bibnamefont {Khudik}}, \bibinfo {author} {\bibfnamefont
  {C.}~\bibnamefont {Siemon}}, \ and\ \bibinfo {author} {\bibfnamefont
  {G.}~\bibnamefont {Shvets}},\ }\bibfield  {title} {\enquote {\bibinfo {title}
  {Phase velocity and particle injection in a self-modulated proton-driven
  plasma wakefield accelerator},}\ }\href {\doibase
  10.1103/PhysRevLett.107.145003} {\bibfield  {journal} {\bibinfo  {journal}
  {Physical Review Letters}\ }\textbf {\bibinfo {volume} {107}},\ \bibinfo
  {pages} {145003} (\bibinfo {year} {2011})}\BibitemShut {NoStop}%
\bibitem [{\citenamefont {Schroeder}\ \emph {et~al.}(2011)\citenamefont
  {Schroeder}, \citenamefont {Benedetti}, \citenamefont {Esarey}, \citenamefont
  {Gr\"uner},\ and\ \citenamefont {Leemans}}]{SCH:GROWTH}%
  \BibitemOpen
  \bibfield  {author} {\bibinfo {author} {\bibfnamefont {C.~B.}\ \bibnamefont
  {Schroeder}}, \bibinfo {author} {\bibfnamefont {C.}~\bibnamefont
  {Benedetti}}, \bibinfo {author} {\bibfnamefont {E.}~\bibnamefont {Esarey}},
  \bibinfo {author} {\bibfnamefont {F.~J.}\ \bibnamefont {Gr\"uner}}, \ and\
  \bibinfo {author} {\bibfnamefont {W.~P.}\ \bibnamefont {Leemans}},\
  }\bibfield  {title} {\enquote {\bibinfo {title} {Growth and phase velocity of
  self-modulated beam-driven plasma waves},}\ }\href {\doibase
  10.1103/PhysRevLett.107.145002} {\bibfield  {journal} {\bibinfo  {journal}
  {Physical Review Letters}\ }\textbf {\bibinfo {volume} {107}},\ \bibinfo
  {pages} {145002} (\bibinfo {year} {2011})}\BibitemShut {NoStop}%
\bibitem [{\citenamefont {{{AWAKE Collaboration}}}(2018)}]{AW:NATURE}%
  \BibitemOpen
  \bibfield  {author} {\bibinfo {author} {\bibnamefont {{{AWAKE
  Collaboration}}}},\ }\bibfield  {title} {\enquote {\bibinfo {title}
  {{Acceleration of electrons in the plasma wakefield of a proton bunch}},}\
  }\href {\doibase 10.1038/s41586-018-0485-4} {\bibfield  {journal} {\bibinfo
  {journal} {Nature}\ }\textbf {\bibinfo {volume} {561}},\ \bibinfo {pages}
  {363--367} (\bibinfo {year} {2018})}\BibitemShut {NoStop}%
\bibitem [{\citenamefont {{L. Verra et al.}}(2022)}]{LIVIO:PRL}%
  \BibitemOpen
  \bibfield  {author} {\bibinfo {author} {\bibnamefont {{L. Verra et al.}}}
  (\bibinfo {collaboration} {AWAKE Collaboration}),\ }\bibfield  {title}
  {\enquote {\bibinfo {title} {Controlled growth of the self-modulation of a
  relativistic proton bunch in plasma},}\ }\href {\doibase
  10.1103/PhysRevLett.129.024802} {\bibfield  {journal} {\bibinfo  {journal}
  {Physical Review Letters}\ }\textbf {\bibinfo {volume} {129}},\ \bibinfo
  {pages} {024802} (\bibinfo {year} {2022})}\BibitemShut {NoStop}%
\bibitem [{\citenamefont {Fang}\ \emph {et~al.}(2014)\citenamefont {Fang},
  \citenamefont {Yakimenko}, \citenamefont {Babzien}, \citenamefont {Fedurin},
  \citenamefont {Kusche}, \citenamefont {Malone}, \citenamefont {Vieira},
  \citenamefont {Mori},\ and\ \citenamefont
  {Muggli}}]{FANG:PhysRevLett.112.045001}%
  \BibitemOpen
  \bibfield  {author} {\bibinfo {author} {\bibfnamefont {Y.}~\bibnamefont
  {Fang}}, \bibinfo {author} {\bibfnamefont {V.~E.}\ \bibnamefont {Yakimenko}},
  \bibinfo {author} {\bibfnamefont {M.}~\bibnamefont {Babzien}}, \bibinfo
  {author} {\bibfnamefont {M.}~\bibnamefont {Fedurin}}, \bibinfo {author}
  {\bibfnamefont {K.~P.}\ \bibnamefont {Kusche}}, \bibinfo {author}
  {\bibfnamefont {R.}~\bibnamefont {Malone}}, \bibinfo {author} {\bibfnamefont
  {J.}~\bibnamefont {Vieira}}, \bibinfo {author} {\bibfnamefont {W.~B.}\
  \bibnamefont {Mori}}, \ and\ \bibinfo {author} {\bibfnamefont
  {P.}~\bibnamefont {Muggli}},\ }\bibfield  {title} {\enquote {\bibinfo {title}
  {Seeding of self-modulation instability of a long electron bunch in a
  plasma},}\ }\href {\doibase 10.1103/PhysRevLett.112.045001} {\bibfield
  {journal} {\bibinfo  {journal} {Physical Review Letters}\ }\textbf {\bibinfo
  {volume} {112}},\ \bibinfo {pages} {045001} (\bibinfo {year}
  {2014})}\BibitemShut {NoStop}%
\bibitem [{\citenamefont {{P. Muggli et al.}}(2017)}]{PATRIC:READINESS}%
  \BibitemOpen
  \bibfield  {author} {\bibinfo {author} {\bibnamefont {{P. Muggli et al.}}}
  (\bibinfo {collaboration} {{AWAKE Collaboration}}),\ }\bibfield  {title}
  {\enquote {\bibinfo {title} {{AWAKE} readiness for the study of the seeded
  self-modulation of a 400 {GeV} proton bunch},}\ }\href {\doibase
  10.1088/1361-6587/aa941c} {\bibfield  {journal} {\bibinfo  {journal} {Plasma
  Physics and Controlled Fusion}\ }\textbf {\bibinfo {volume} {60}},\ \bibinfo
  {pages} {014046} (\bibinfo {year} {2017})}\BibitemShut {NoStop}%
\bibitem [{\citenamefont {Keinigs}\ and\ \citenamefont
  {Jones}(1987)}]{JONES:LIN_THEORY}%
  \BibitemOpen
  \bibfield  {author} {\bibinfo {author} {\bibfnamefont {R.}~\bibnamefont
  {Keinigs}}\ and\ \bibinfo {author} {\bibfnamefont {M.~E.}\ \bibnamefont
  {Jones}},\ }\bibfield  {title} {\enquote {\bibinfo {title} {Two‐dimensional
  dynamics of the plasma wakefield accelerator},}\ }\href {\doibase
  10.1063/1.866183} {\bibfield  {journal} {\bibinfo  {journal} {The Physics of
  Fluids}\ }\textbf {\bibinfo {volume} {30}},\ \bibinfo {pages} {252--263}
  (\bibinfo {year} {1987})}\BibitemShut {NoStop}%
\bibitem [{\citenamefont {{G.~Demeter et al.}}(2021)}]{GABOR:PLASMA}%
  \BibitemOpen
  \bibfield  {author} {\bibinfo {author} {\bibnamefont {{G.~Demeter et al.}}},\
  }\bibfield  {title} {\enquote {\bibinfo {title} {Long-range propagation of
  ultrafast ionizing laser pulses in a resonant nonlinear medium},}\ }\href
  {\doibase 10.1103/PhysRevA.104.033506} {\bibfield  {journal} {\bibinfo
  {journal} {Physical Review A}\ }\textbf {\bibinfo {volume} {104}},\ \bibinfo
  {pages} {033506} (\bibinfo {year} {2021})}\BibitemShut {NoStop}%
\bibitem [{\citenamefont {Plyushchev}\ \emph {et~al.}(2017)\citenamefont
  {Plyushchev}, \citenamefont {Kersevan}, \citenamefont {Petrenko},\ and\
  \citenamefont {Muggli}}]{SOURCE:Plyushchev_2018}%
  \BibitemOpen
  \bibfield  {author} {\bibinfo {author} {\bibfnamefont {G.}~\bibnamefont
  {Plyushchev}}, \bibinfo {author} {\bibfnamefont {R.}~\bibnamefont
  {Kersevan}}, \bibinfo {author} {\bibfnamefont {A.}~\bibnamefont {Petrenko}},
  \ and\ \bibinfo {author} {\bibfnamefont {P.}~\bibnamefont {Muggli}},\
  }\bibfield  {title} {\enquote {\bibinfo {title} {A rubidium vapor source for
  a plasma source for awake},}\ }\href {\doibase 10.1088/1361-6463/aa9dd7}
  {\bibfield  {journal} {\bibinfo  {journal} {Journal of Physics D: Applied
  Physics}\ }\textbf {\bibinfo {volume} {51}},\ \bibinfo {pages} {025203}
  (\bibinfo {year} {2017})}\BibitemShut {NoStop}%
\bibitem [{\citenamefont {{J.S. Schmidt et al.}}(2017)}]{SCHMIDT:PROTON_LINE}%
  \BibitemOpen
  \bibfield  {author} {\bibinfo {author} {\bibnamefont {{J.S. Schmidt et
  al.}}},\ }\bibfield  {title} {\enquote {\bibinfo {title} {{AWAKE} {P}roton
  {B}eam {C}ommissioning},}\ }\bibfield  {booktitle} {\emph {\bibinfo
  {booktitle} {Proc. IPAC'17}},\ }\href {\doibase
  https://doi.org/10.18429/JACoW-IPAC2017-TUPIK032} {\ \bibinfo {series}
  {International Particle Accelerator Conference},\ \bibinfo {eid} {TUPIK032}
  (\bibinfo {year} {2017})},\ \bibinfo {note}
  {https://doi.org/10.18429/JACoW-IPAC2017-TUPIK032}\BibitemShut {NoStop}%
\bibitem [{\citenamefont {{T. Nechaeva, L. Verra, G. Zevi Della Porta and P.
  Muggli}}(2023)}]{Nechaeva_2023}%
  \BibitemOpen
  \bibfield  {author} {\bibinfo {author} {\bibnamefont {{T. Nechaeva, L. Verra,
  G. Zevi Della Porta and P. Muggli}}},\ }\bibfield  {title} {\enquote
  {\bibinfo {title} {{A method for obtaining 3D charge density distribution of
  a self-modulated proton bunch}},}\ }\href {\doibase
  10.1088/1742-6596/2420/1/012063} {\bibfield  {journal} {\bibinfo  {journal}
  {Journal of Physics: Conference Series}\ }\textbf {\bibinfo {volume}
  {2420}},\ \bibinfo {pages} {012063} (\bibinfo {year} {2023})}\BibitemShut
  {NoStop}%
\bibitem [{unc()}]{uncertainties}%
  \BibitemOpen
  \href@noop {} {\bibinfo  {journal} {The uncertainty on each parameter of the
  fits is obtained from the covariance matrix. For all cases shown in
  Fig.~\ref{fig:2}(c), the uncertainty on $\sigma_{x,y}^*$ is $<12\,\mu$m, on
  $\epsilon_{x,y}$ is $<0.2\,$mm-mrad, on $z_{x,y}^*$ is $<0.8\,$m}\
  }\BibitemShut {NoStop}%
\bibitem [{\citenamefont {{{M. Turner, P. Muggli et
  al.}}}(2020)}]{MARLENE:PRAB}%
  \BibitemOpen
\bibfield  {journal} {  }\bibfield  {author} {\bibinfo {author} {\bibnamefont
  {{{M. Turner, P. Muggli et al.}}}} (\bibinfo {collaboration} {AWAKE
  Collaboration}),\ }\bibfield  {title} {\enquote {\bibinfo {title}
  {Experimental study of wakefields driven by a self-modulating proton bunch in
  plasma},}\ }\href {\doibase 10.1103/PhysRevAccelBeams.23.081302} {\bibfield
  {journal} {\bibinfo  {journal} {Physical Review Accelerators and Beams}\
  }\textbf {\bibinfo {volume} {23}},\ \bibinfo {pages} {081302} (\bibinfo
  {year} {2020})}\BibitemShut {NoStop}%
\bibitem [{\citenamefont {Bachmann}\ and\ \citenamefont
  {Muggli}(2020)}]{Bachmann_2020}%
  \BibitemOpen
  \bibfield  {author} {\bibinfo {author} {\bibfnamefont {A.-M.}\ \bibnamefont
  {Bachmann}}\ and\ \bibinfo {author} {\bibfnamefont {P.}~\bibnamefont
  {Muggli}},\ }\bibfield  {title} {\enquote {\bibinfo {title} {Determination of
  the charge per micro-bunch of a self-modulated proton bunch using a streak
  camera},}\ }\href {\doibase 10.1088/1742-6596/1596/1/012005} {\bibfield
  {journal} {\bibinfo  {journal} {Journal of Physics: Conference Series}\
  }\textbf {\bibinfo {volume} {1596}},\ \bibinfo {pages} {012005} (\bibinfo
  {year} {2020})}\BibitemShut {NoStop}%
\bibitem [{\citenamefont {Batsch}(2020)}]{FABIAN:MARKER}%
  \BibitemOpen
  \bibfield  {author} {\bibinfo {author} {\bibfnamefont {F.}~\bibnamefont
  {Batsch}},\ }\bibfield  {title} {\enquote {\bibinfo {title} {Setup and
  characteristics of a timing reference signal with sub-ps accuracy for
  {AWAKE}},}\ }\href {\doibase 10.1088/1742-6596/1596/1/012006} {\bibfield
  {journal} {\bibinfo  {journal} {Journal of Physics: Conference Series}\
  }\textbf {\bibinfo {volume} {1596}},\ \bibinfo {pages} {012006} (\bibinfo
  {year} {2020})}\BibitemShut {NoStop}%
\bibitem [{\citenamefont {{K. Rieger, A. Caldwell, O. Reimann, R. Tarkeshian
  and P. Muggli}}(2017)}]{KARL:STREAK}%
  \BibitemOpen
  \bibfield  {author} {\bibinfo {author} {\bibnamefont {{K. Rieger, A.
  Caldwell, O. Reimann, R. Tarkeshian and P. Muggli}}},\ }\bibfield  {title}
  {\enquote {\bibinfo {title} {{GHz modulation detection using a streak camera:
  Suitability of streak cameras in the AWAKE experiment}},}\ }\href {\doibase
  10.1063/1.4975380} {\bibfield  {journal} {\bibinfo  {journal} {Review of
  Scientific Instruments}\ }\textbf {\bibinfo {volume} {88}},\ \bibinfo {pages}
  {025110} (\bibinfo {year} {2017})},\ \Eprint
  {http://arxiv.org/abs/https://doi.org/10.1063/1.4975380}
  {https://doi.org/10.1063/1.4975380} \BibitemShut {NoStop}%
\bibitem [{\citenamefont {Lee}\ and\ \citenamefont
  {Lampe}(1973)}]{CFI:PhysRevLett.31.1390}%
  \BibitemOpen
  \bibfield  {author} {\bibinfo {author} {\bibfnamefont {R.}~\bibnamefont
  {Lee}}\ and\ \bibinfo {author} {\bibfnamefont {M.}~\bibnamefont {Lampe}},\
  }\bibfield  {title} {\enquote {\bibinfo {title} {Electromagnetic
  instabilities, filamentation, and focusing of relativistic electron beams},}\
  }\href {\doibase 10.1103/PhysRevLett.31.1390} {\bibfield  {journal} {\bibinfo
   {journal} {Physical Review Letters}\ }\textbf {\bibinfo {volume} {31}},\
  \bibinfo {pages} {1390--1393} (\bibinfo {year} {1973})}\BibitemShut {NoStop}%
\bibitem [{\citenamefont {Allen}\ \emph {et~al.}(2012)\citenamefont {Allen},
  \citenamefont {Yakimenko}, \citenamefont {Babzien}, \citenamefont {Fedurin},
  \citenamefont {Kusche},\ and\ \citenamefont
  {Muggli}}]{CFI:PhysRevLett.109.185007}%
  \BibitemOpen
  \bibfield  {author} {\bibinfo {author} {\bibfnamefont {B.}~\bibnamefont
  {Allen}}, \bibinfo {author} {\bibfnamefont {V.}~\bibnamefont {Yakimenko}},
  \bibinfo {author} {\bibfnamefont {M.}~\bibnamefont {Babzien}}, \bibinfo
  {author} {\bibfnamefont {M.}~\bibnamefont {Fedurin}}, \bibinfo {author}
  {\bibfnamefont {K.}~\bibnamefont {Kusche}}, \ and\ \bibinfo {author}
  {\bibfnamefont {P.}~\bibnamefont {Muggli}},\ }\bibfield  {title} {\enquote
  {\bibinfo {title} {Experimental study of current filamentation
  instability},}\ }\href {\doibase 10.1103/PhysRevLett.109.185007} {\bibfield
  {journal} {\bibinfo  {journal} {Physical Review Letters}\ }\textbf {\bibinfo
  {volume} {109}},\ \bibinfo {pages} {185007} (\bibinfo {year}
  {2012})}\BibitemShut {NoStop}%
\bibitem [{\citenamefont {{P. Muggli, L. Verra, N. Lopes, A.
  Sublet}}()}]{PATRIC:IPAC23}%
  \BibitemOpen
  \bibfield  {author} {\bibinfo {author} {\bibnamefont {{P. Muggli, L. Verra,
  N. Lopes, A. Sublet}}},\ }\bibfield  {title} {\enquote {\bibinfo {title}
  {{Self-modulation and current filamentation instabilities of long and wide
  proton bunches in plasma}},}\ }\bibfield  {booktitle} {\emph {\bibinfo
  {booktitle} {Contribution to IPAC '23}},\ }\href@noop {} {\ }\BibitemShut
  {NoStop}%
\bibitem [{\citenamefont {Rosenzweig}(1987)}]{BLOWOUT:ROSENZ}%
  \BibitemOpen
  \bibfield  {author} {\bibinfo {author} {\bibfnamefont {J.~B.}\ \bibnamefont
  {Rosenzweig}},\ }\bibfield  {title} {\enquote {\bibinfo {title} {Nonlinear
  plasma dynamics in the plasma wake-field accelerator},}\ }\href {\doibase
  10.1103/PhysRevLett.58.555} {\bibfield  {journal} {\bibinfo  {journal}
  {Physical Review Letters}\ }\textbf {\bibinfo {volume} {58}},\ \bibinfo
  {pages} {555--558} (\bibinfo {year} {1987})}\BibitemShut {NoStop}%
\bibitem [{\citenamefont {Muggli}\ \emph {et~al.}(2004)\citenamefont {Muggli},
  \citenamefont {Blue}, \citenamefont {Clayton}, \citenamefont {Deng},
  \citenamefont {Decker}, \citenamefont {Hogan}, \citenamefont {Huang},
  \citenamefont {Iverson}, \citenamefont {Joshi}, \citenamefont {Katsouleas},
  \citenamefont {Lee}, \citenamefont {Lu}, \citenamefont {Marsh}, \citenamefont
  {Mori}, \citenamefont {O'Connell}, \citenamefont {Raimondi}, \citenamefont
  {Siemann},\ and\ \citenamefont {Walz}}]{PATRIC:MATCHING}%
  \BibitemOpen
  \bibfield  {author} {\bibinfo {author} {\bibfnamefont {P.}~\bibnamefont
  {Muggli}}, \bibinfo {author} {\bibfnamefont {B.~E.}\ \bibnamefont {Blue}},
  \bibinfo {author} {\bibfnamefont {C.~E.}\ \bibnamefont {Clayton}}, \bibinfo
  {author} {\bibfnamefont {S.}~\bibnamefont {Deng}}, \bibinfo {author}
  {\bibfnamefont {F.-J.}\ \bibnamefont {Decker}}, \bibinfo {author}
  {\bibfnamefont {M.~J.}\ \bibnamefont {Hogan}}, \bibinfo {author}
  {\bibfnamefont {C.}~\bibnamefont {Huang}}, \bibinfo {author} {\bibfnamefont
  {R.}~\bibnamefont {Iverson}}, \bibinfo {author} {\bibfnamefont
  {C.}~\bibnamefont {Joshi}}, \bibinfo {author} {\bibfnamefont {T.~C.}\
  \bibnamefont {Katsouleas}}, \bibinfo {author} {\bibfnamefont
  {S.}~\bibnamefont {Lee}}, \bibinfo {author} {\bibfnamefont {W.}~\bibnamefont
  {Lu}}, \bibinfo {author} {\bibfnamefont {K.~A.}\ \bibnamefont {Marsh}},
  \bibinfo {author} {\bibfnamefont {W.~B.}\ \bibnamefont {Mori}}, \bibinfo
  {author} {\bibfnamefont {C.~L.}\ \bibnamefont {O'Connell}}, \bibinfo {author}
  {\bibfnamefont {P.}~\bibnamefont {Raimondi}}, \bibinfo {author}
  {\bibfnamefont {R.}~\bibnamefont {Siemann}}, \ and\ \bibinfo {author}
  {\bibfnamefont {D.}~\bibnamefont {Walz}},\ }\bibfield  {title} {\enquote
  {\bibinfo {title} {Meter-scale plasma-wakefield accelerator driven by a
  matched electron beam},}\ }\href {\doibase 10.1103/PhysRevLett.93.014802}
  {\bibfield  {journal} {\bibinfo  {journal} {Physical Review Letters}\
  }\textbf {\bibinfo {volume} {93}},\ \bibinfo {pages} {014802} (\bibinfo
  {year} {2004})}\BibitemShut {NoStop}%
\bibitem [{\citenamefont {Gschwendtner}(2021)}]{EDDA:IPAC}%
  \BibitemOpen
  \bibfield  {author} {\bibinfo {author} {\bibfnamefont {E.}~\bibnamefont
  {Gschwendtner}},\ }\bibfield  {title} {\enquote {\bibinfo {title} {{Awake Run
  2 at CERN}},}\ }\bibfield  {booktitle} {\emph {\bibinfo {booktitle} {Proc.
  IPAC'21}},\ }\href {\doibase 10.18429/JACoW-IPAC2021-TUPAB159} {\ \bibinfo
  {series} {International Particle Accelerator Conference},\ \bibinfo {eid}
  {TUPAB159} (\bibinfo {year} {2021})},\ \bibinfo {note}
  {https://doi.org/10.18429/JACoW-IPAC2021-TUPAB159}\BibitemShut {NoStop}%
\bibitem [{\citenamefont {Muggli}(2020)}]{PATRIC:EAAC}%
  \BibitemOpen
  \bibfield  {author} {\bibinfo {author} {\bibfnamefont {P.}~\bibnamefont
  {Muggli}},\ }\bibfield  {title} {\enquote {\bibinfo {title} {{Physics to plan
  {AWAKE} Run 2}},}\ }\href {\doibase 10.1088/1742-6596/1596/1/012008}
  {\bibfield  {journal} {\bibinfo  {journal} {Journal of Physics: Conference
  Series}\ }\textbf {\bibinfo {volume} {1596}},\ \bibinfo {pages} {012008}
  (\bibinfo {year} {2020})}\BibitemShut {NoStop}%
\bibitem [{\citenamefont {Verra}, \citenamefont {Gschwendtner},\ and\
  \citenamefont {Muggli}(2020)}]{LIVIO:EAAC}%
  \BibitemOpen
  \bibfield  {author} {\bibinfo {author} {\bibfnamefont {L.}~\bibnamefont
  {Verra}}, \bibinfo {author} {\bibfnamefont {E.}~\bibnamefont {Gschwendtner}},
  \ and\ \bibinfo {author} {\bibfnamefont {P.}~\bibnamefont {Muggli}},\
  }\bibfield  {title} {\enquote {\bibinfo {title} {{Study of external electron
  beam injection into proton driven plasma wakefields for AWAKE Run 2}},}\
  }\href {\doibase 10.1088/1742-6596/1596/1/012007} {\bibfield  {journal}
  {\bibinfo  {journal} {Journal of Physics: Conference Series}\ }\textbf
  {\bibinfo {volume} {1596}},\ \bibinfo {pages} {012007} (\bibinfo {year}
  {2020})}\BibitemShut {NoStop}%
\bibitem [{\citenamefont {Baistrukov}\ and\ \citenamefont
  {Lotov}(2022)}]{Baistrukov_2022}%
  \BibitemOpen
  \bibfield  {author} {\bibinfo {author} {\bibfnamefont {M.~A.}\ \bibnamefont
  {Baistrukov}}\ and\ \bibinfo {author} {\bibfnamefont {K.~V.}\ \bibnamefont
  {Lotov}},\ }\bibfield  {title} {\enquote {\bibinfo {title} {Evolution of
  equilibrium particle beams in plasma under external wakefields},}\ }\href
  {\doibase 10.1088/1361-6587/ac6ffe} {\bibfield  {journal} {\bibinfo
  {journal} {Plasma Physics and Controlled Fusion}\ }\textbf {\bibinfo {volume}
  {64}},\ \bibinfo {pages} {075003} (\bibinfo {year} {2022})}\BibitemShut
  {NoStop}%
\end{thebibliography}%


%merlin.mbs aipnum4-1.bst 2010-07-25 4.21a (PWD, AO, DPC) hacked
%Control: key (0)
%Control: author (8) initials jnrlst
%Control: editor formatted (1) identically to author
%Control: production of article title (0) allowed
%Control: page (1) range
%Control: year (1) truncated
%Control: production of eprint (0) enabled
%

\end{document}